\documentclass[lettersize,journal]{IEEEtran}
\usepackage{amsmath,amsfonts}
\usepackage{algorithm}
\usepackage{algorithmic}
\usepackage{array}
\usepackage[caption=false,font=normalsize,labelfont=sf,textfont=sf]{subfig}
\usepackage{textcomp}
\usepackage{stfloats}
\usepackage{url}
\usepackage{verbatim}
\usepackage{graphicx}
\usepackage{cite}
\usepackage{color}
\begin{document}

\title{Precoding-free Hierarchical Rate-Splitting Multiple Access via Stacked Intelligent Metasurface}

\author{
    Hiroaki~Hashida,~\IEEEmembership{Member,~IEEE},
    Boya~Di,~\IEEEmembership{Senior Member,~IEEE}% <-this % stops space.
\thanks{H. Hashida is with Frontier Research Institute for Interdisciplinary Sciences, Tohoku University, Sendai, Japan.
E-mail: hiroaki.hashida.d6@tohoku.ac.jp.}% <-this % stops a space

\thanks{Boya Di is State Key Laboratory of Advanced Optical Communication Systems and Networks, School of Electronics, Peking University, Beijing, China,
E-mail: diboya@pku.edu.cn.}% <-this % stops a space
}

\markboth{Submitted to IEEE Internet of Things Journal}%
{Shell \MakeLowercase{\textit{et al.}}: A Sample Article Using IEEEtran.cls for IEEE Journals}

\maketitle

\begin{abstract}
Interference management is a central bottleneck in dense multi-antenna wireless networks. Therefore, in this study, we present a digital precoding-free hierarchical rate-splitting multiple access (HRSMA) architecture assisted by a stacked intelligent metasurface (SIM) to achieve high spectral efficiency and user fairness with reduced hardware complexity. 
In the proposed system, the base station performs only scalar power allocation, while a multi-layer SIM acts as a wave-domain processor that spatially separates users and mitigates interference via nonlinear wavefront reconfiguration. 
This design eliminates the need for digital or hybrid precoding, drastically reducing the baseband computations.
A joint optimization problem is formulated to maximize the minimum user rate by jointly optimizing SIM phase shifts, power allocation, and user grouping. 
To efficiently solve the resulting non-convex problem, an alternating optimization algorithm is developed, combining simultaneous perturbation stochastic approximation (SPSA) for SIM configuration and power control with clustering-based grouping refinement. 
Simulation results demonstrate that the proposed SIM-aided HRSMA achieves substantial gains in both spectral efficiency and fairness compared to hybrid beamforming and non-precoding baselines. 
Specifically, SIM-aided HRSMA attains comparable or superior minimum rates with significantly fewer active antennas by exploiting the additional wave-domain degrees of freedom provided by multi-layer SIMs. 
These findings highlight the potential of SIM-aided HRSMA as a low-cost, energy-efficient, and scalable solution for beyond-6G networks.
\end{abstract}

\begin{IEEEkeywords}
Stacked intelligent metasurface (SIM), hierarchical rate-splitting multiple access (HRSMA), rate-splitting multiple access (RSMA), wave-domain beamforming, simultaneous perturbation stochastic approximation (SPSA), max-min optimization.
\end{IEEEkeywords}

\section{Introduction}

The sixth generation (6G) mobile communication system is envisioned not merely as an extension of the capabilities of 5G networks but as a foundation for realizing the internet of everything (IoE)~\cite{IoT,IoTJ_1,mina}. Such a system is expected to guarantee ultra-high data rates, massive connectivity, extreme reliability, and heterogeneous quality of service (QoS). To achieve these ambitious goals, 6G networks must efficiently utilize wireless resources and adopt more powerful interference management strategies~\cite{6G_resource_management,Boya_HDMA_JSAC}. Interference management remains a central bottleneck in dense multi-antenna wireless networks. Classical multiple access schemes, such as orthogonal multiple access (OMA) and non-orthogonal multiple access (NOMA), represent two contrasting design philosophies~\cite{OMA_NOMA}. OMA completely avoids inter-user interference by allocating orthogonal time, frequency, or spatial resources at the cost of reduced spectral efficiency. In contrast, NOMA improves spectral efficiency through resource sharing among users; however, it inevitably introduces inter-user interference, which increases receiver complexity and may lead to performance degradation~\cite{NOMA_1,NOMA_2}. Space-division multiple access (SDMA) employing linear precoding provides an alternative by exploiting spatial multiplexing. However, its interference management capability relies heavily on the orthogonality among user channels and requires substantial digital signal processing resources.

\begin{figure*}[t]
	\centering
	\includegraphics[width=1.0\hsize]{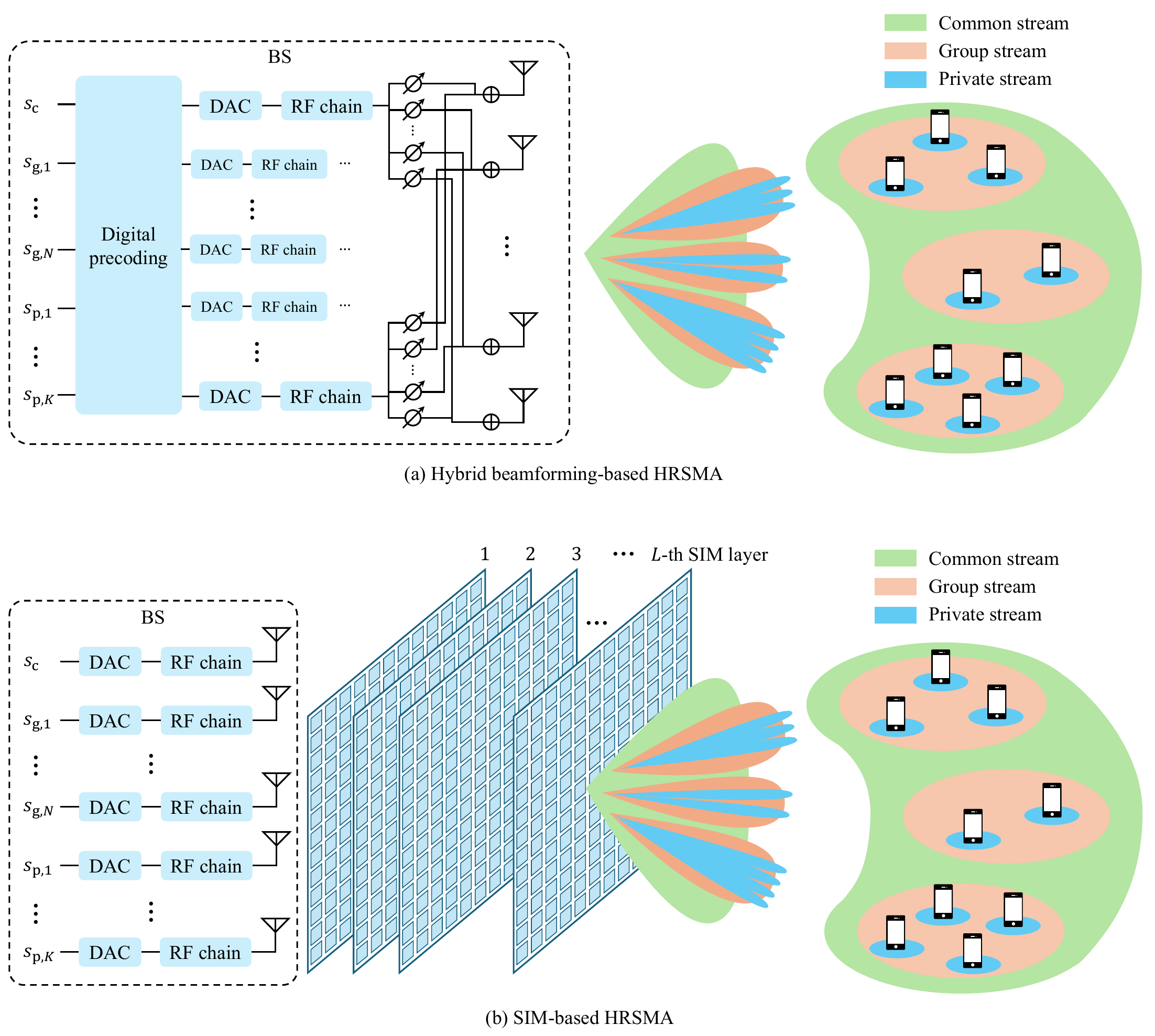}
	\caption{Comparison of conventional hybrid beamforming-based HRSMA and SIM-aided HRSMA.}
	\label{fig:SIMandHBF_HRSMA}
\end{figure*}

To address these limitations, rate-splitting multiple access (RSMA)~\cite{RSMA_2,RSMA_4} and its hierarchical extension, hierarchical RSMA (HRSMA), also referred to as grouping-based RSMA (G-RSMA)~\cite{HRSMA_1,HRSMA_2,HRSMA_3}, have recently attracted increasing attention. RSMA introduces a flexible framework that splits each user's message into common and private parts, allowing partial interference decoding and partial interference treating-as-noise. This enables a smooth transition between OMA, SDMA, and NOMA, achieving robust performance across a wide range of channel conditions and user deployments. HRSMA further extends this principle by introducing multiple layers of common streams, which improves fairness among users in dense multi-user environments.
However, both RSMA and HRSMA rely on complex digital precoding at the transmitter, which requires advanced hardware configurations. In particular, implementing digital beamforming requires an independent radiofrequency (RF) chain and high-speed analog-to-digital and digital-to-analog converters for each antenna element, which significantly increases hardware cost, power consumption, and system implementation complexity~\cite{RSMA_1}.
To mitigate these issues, hybrid beamforming (HBF), which combines analog and digital beamforming, has been proposed~\cite{RSMA_beamforming}. Nevertheless, it still requires a large number of active antenna elements and phase shifters, which is a major challenge for implementing scalable large-scale antenna systems as illustrated in Fig.~\ref{fig:SIMandHBF_HRSMA}~(a).

In contrast, stacked intelligent metasurfaces (SIMs) have recently emerged as a promising hardware technology for realizing highly flexible beamforming directly in the wave domain. 
Unlike conventional single-layer intelligent reflecting surfaces (IRSs) that only impose passive phase shifts on incident waves~\cite{Boya_IRS_TVT,NZhang_1,Hashida_PartialBlockage_TCCN}, SIMs comprise multiple transmissive metasurface layers that perform cascaded transformations of electromagnetic waves~\cite{SIM_1}. 
SIMs can achieve complex wave-based signal processing comparable to digital beamforming by jointly optimizing the transmission coefficients of the layers with significantly lower power consumption and hardware complexity. Thus, they are a compelling alternative or complement to digital beamforming architectures for future 6G networks.

These two lines, RSMA/HRSMA and SIM, are naturally synergistic. The layered common/private architecture maps well to the multi-layer physical degrees of freedom of SIMs, suggesting a new class of transmitters in which hierarchical RSMA is realized with minimal or no digital spatial precoding, while the SIM handles inter-user separation and wavefront shaping in the physical domain.

\subsection{Prior Works}
\subsubsection{RSMA and HRSMA}
Extensive research has been devoted to RSMA and HRSMA, as promising multiple access frameworks for beyond 5G and 6G systems~\cite{RSMA_1,RSMA_4,HRSMA_1,HRSMA_3}.
Early studies established RSMA as a general strategy that unifies and outperforms conventional schemes such as OMA, SDMA, and NOMA by adaptively controlling the degree of interference decoding and treating-as-noise~\cite{RSMA_3}. 
Pioneering works have demonstrated that RSMA yields superior spectral and energy efficiency for heterogeneous user deployments, and dynamic traffic loads~\cite{RSMA_2}. 
Subsequently, HRSMA was introduced to further enhance system robustness by enabling multi-layer common streams decoded by user subsets~\cite{HRSMA_2}. 
This hierarchical message structure facilitates finer control over interference distribution and fairness optimization. However, these approaches generally assume full-digital precoding at the base station, resulting in high computational complexity and hardware cost. 
Although recent studies have explored low-complexity designs and hybrid precoding for RSMA, the digital domain implementation remains a significant bottleneck for the large-scale antenna systems envisioned in 6G. 

\subsubsection{IRS-enhanced multiple access}
IRSs have emerged as a transformative technology that redefines the way electromagnetic waves are manipulated~\cite{Hashida_FluidAntenna_JSAC,IRS_1,NZhang_2,Hibi_near_TCCN}. 
Numerous studies have investigated the integration of IRSs with traditional multiple access schemes. 
For instance, IRS-aided NOMA systems have been shown to improve spectral efficiency through controllable reflections that enhance weak users channels~\cite{Boya_IRSNOMA_NetworkM,IRS_2,Hashida_sharing_WCM,IRS_3}. 
Similarly, IRS-assisted SDMA and OMA have been analyzed for improved spatial multiplexing and interference mitigation~\cite{Hashida_Sharing_TCCN}. 
Moreover, the combination of IRS and RSMA has recently been investigated to exploit the complementary strengths of both paradigms. 
In these works, the IRS is employed to enhance the cascaded base station (BS)-IRS-user channels, while RSMA/HRSMA provides a robust transmission framework that is resilient to residual interference and imperfect channel estimation~\cite{IRS_RSMA_1,IRS_RSMA_2,IRS_RSMA_3,IRS_HRSMA_1,IRS_HRSMA_2,IRS_HRSMA_3}. 
Simulation results have confirmed that IRS-assisted RSMA and HRSMA can outperform IRS-aided NOMA and SDMA systems in terms of spectral and energy efficiency. 
Nonetheless, these studies typically assume conventional single-layer IRSs and retain digital precoding at the base station, which limits the potential to fully offload spatial processing to the physical layer.
However, conventional single-layer IRSs function primarily as passive reflectors and are fundamentally constrained by the double path-loss effect and limited spatial processing capability, which restricts their performance in highly dynamic multi-user scenarios~\cite{Hibi_Near_WCM}. 

\subsubsection{SIM-enhanced multiple access}
To overcome these limitations, SIMs have been proposed as an evolution of RIS technology.
Recent works have demonstrated that SIMs can achieve high-precision focusing and multi-user beam shaping while substantially reducing the need for active RF chains. 
Several works have explored SIM architectures in multiple-input multiple-output (MIMO) systems to increase the number of spatial streams by exploiting their wave-domain degrees of freedom~\cite{SIM_1,SIM_2}. 
Moreover, leveraging the intrinsic cascaded structure of SIMs, some researchers have attempted to emulate neural network operations in the electromagnetic domain, enabling analog implementations of nonlinear signal transformations~\cite{SIM_machine_learning}. 
Overall, despite rapid progress in SIM-enabled physical-layer technologies, most studies focus on holographic MIMO and near-field beam focusing, with limited efforts toward integrating SIM with multiple access schemes~\cite{SIM_RSMA}.
To the best of our knowledge, no existing work has investigated the integration of HRSMA with SIM architectures. This unexplored direction represents a promising opportunity: the layered message structure of HRSMA naturally aligns with the multi-layer physical configuration of SIMs, suggesting a synergistic framework in which message hierarchy and spatial wave transformation are co-optimized. Such integration could enable a fully precoding-free hierarchical multiple access paradigm, achieving interference management, user fairness, and adaptive wavefront shaping directly in the physical domain without relying on complex digital processing.

\subsection{Contributions}
In this report, we investigate a downlink SIM-aided HRSMA system to achieve high spectral efficiency and scalability with minimal hardware complexity, as shown in Fig.~\ref{fig:SIMandHBF_HRSMA} (b).
In the proposed architecture, each transmit antenna corresponds to a distinct message stream, where only scalar power allocation is applied, and these messages are directly transmitted without any precoding. The SIM operates as a wave-domain processor, effectively serving as an alternative to conventional digital precoders by performing spatial signal transformation via programmable meta-atom layers.
The main contributions of this investigation are summarized as follows:
\begin{itemize}
    \item A precoding-free HRSMA architecture is proposed where the BS employs only scalar power allocation, while a stacked intelligent metasurface performs wave-domain multiuser separation and common-stream shaping. This architecture targets low-power, low-cost transmitters by removing digital spatial precoding.
    \item A joint optimization problem is formulated that simultaneously determines the layer-wise (common / group / private) power allocation and the configuration of the SIM layers. To ensure multiuser fairness, the objective is to maximize the minimum user rate. The proposed formulation exploits the additional degrees of freedom provided by the SIM to control the transmission channel in the wave domain, thereby reinforcing the interference-management capability of HRSMA systems.
    \item To efficiently solve the formulated non-convex optimization problem arising from coupling between power allocation and SIM phase shifts, as well as the unit-modulus constraints on the SIM phases, an alternating optimization (AO)-based algorithm is developed. The algorithm decomposes the overall problem into several tractable subproblems, enabling scalable and computationally efficient analysis.
    \item Through extensive numerical simulations, a comprehensive comparison is performed between the proposed SIM-based HRSMA scheme and conventional HBF-based HRSMA architectures. The results demonstrate that the proposed architecture, benefiting from the SIM's nonlinear wave-domain beamforming capability, significantly reduces baseband spatial processing complexity while achieving superior performance. Consequently, the proposed approach achieves enhanced max-min fairness with lower hardware cost compared to traditional architectures.
\end{itemize}

\subsection{Paper Organization}
The remainder of this report is organized as follows. 
Section~\ref{section:system_model} introduces the system model of the proposed SIM-aided HRSMA architecture, including the physical structure of the SIM and the corresponding transmission model. 
Section~\ref{section:problem_formulation} formulates the joint optimization problem for SIM phase-shift configuration, power allocation, and user grouping under fairness constraints. 
Section~\ref{section:joint_optimization} presents the proposed AO framework, which decomposes the overall non-convex problem into three tractable subproblems and develops efficient stochastic optimization algorithms for each. 
Section~\ref{section:performance_evaluation} provides numerical simulation results to evaluate the performance of the proposed method in comparison with conventional HBF and non-precoding baselines. 
Section~\ref{section:discussion} discusses the physical insights, design trade-offs, and emphasizes the nonlinear wave-domain beamforming nature of SIM and its synergy with HRSMA.  
Finally, Section~\ref{section:conclusion} summarizes the main conclusions and discusses potential future research directions.

\subsection{Notations}
Scalars, vectors, and matrices are denoted by italic letters $x$, bold lowercase letters $\mathbf{x}$, 
and bold uppercase letters $\mathbf{X}$, respectively. 
The operators $(\cdot)^{\mathrm{T}}$ and $(\cdot)^{\mathrm{H}}$ denote the transpose and Hermitian transpose, respectively. 
$\mathbb{C}^{m\times n}$ denotes the set of complex matrices of size $m\times n$, 
and $\mathcal{CN}(\mu, \sigma^2)$ represents a circularly symmetric complex Gaussian distribution with mean $\mu$ and variance $\sigma^2$. 
The element-wise product is represented by $\odot$.

\begin{figure}[t]
	\centering
	\includegraphics[width=1.0\hsize]{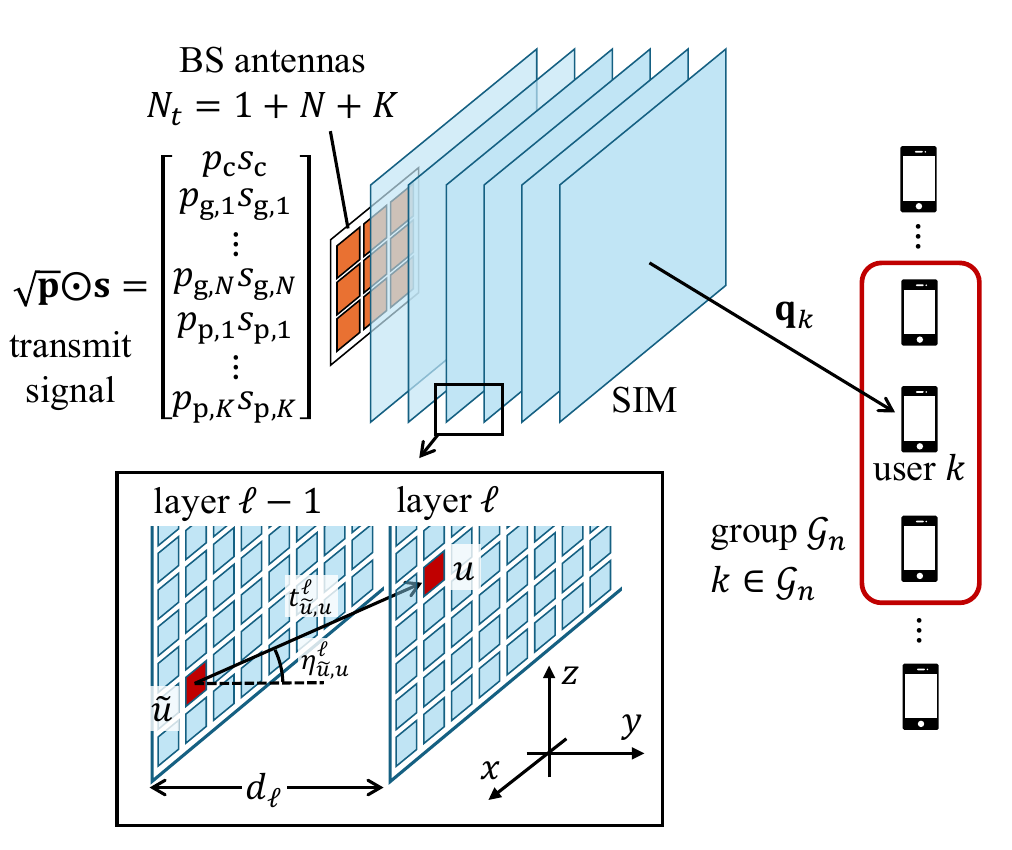}
	\caption{The proposed SIM-aided HRSMA system.}
	\label{fig:system_model}
\end{figure}

\section{System Model}
\label{section:system_model}
In this report, we consider a downlink SIM-aided HRSMA system, where a BS serves $K$ single-antenna users illustrated in Fig.~\ref{fig:system_model}.
We first introduce the architecture of the transmitter equipped with the SIM in Section~\ref{section:sim-assisted_transmitter_model}.
Then, we provide the details of the SIM-aided RSMA transmission framework in Section~\ref{section:sim-aided_rsma_transmission_framework}.

\subsection{SIM-based Transmitter Model}
\label{section:sim-assisted_transmitter_model}
We consider a SIM that is installed inside the radome of the BS. 
All SIM layers and BS antenna array are parallel to the $x$-$z$ plane. 
The BS is equipped with $N_t$ antennas, and the SIM is composed of $L$ transmissive layers, each containing $U$($=U_a \times U_a$) programmable meta-atoms. 
The layer spacing between the SIM layers or between the BS antennas and the $1$-st SIM layer is denoted by $d_{\ell}$.
Let $\mathbf{W}^{\ell} \in \mathbb{C}^{U \times U}$ denote the inter-layer channel matrix between the $(\ell - 1)$-th and $\ell$-th layers of the SIM. 
The $(\tilde{u},u)$-th element of $\mathbf{W}^{\ell}$ represents the complex propagation coefficient between the $\tilde{u}$-th meta-atom of layer $(\ell-1)$ and the $u$-th meta-atom of layer $\ell$, and $\mathbf{W}^{\ell}$ is given by:
\begin{equation}
    \mathbf{W}_{\ell} =
    \begin{bmatrix}
        w^{\ell}_{1,1} & \cdots & w^{\ell}_{1,U} \\
        \vdots & \ddots & \vdots \\
        w^{\ell}_{U,1} & \cdots & w^{\ell}_{U,U}
    \end{bmatrix},
    \quad 
    w^{\ell}_{\tilde{u},u} \in \mathbb{C}.
\end{equation}
Furthermore, all meta-atoms in each SIM layer are modeled as a uniform planar array (UPA). 
The inter-element distance between the $\tilde{u}$-th element of the $(\ell-1)$-th layer and the $u$-th element of the $\ell$-th layer is given by
\begin{equation}
    t_{\tilde{u},u}^{\ell} 
    = \sqrt{
        d_{\ell}^2 
        + (x_{u,\ell} - \tilde{x}_{\tilde{u},\ell-1})^2 
        + (z_{u,\ell} - \tilde{z}_{\tilde{u},\ell-1})^2
    }.
\end{equation}
where $(x_{u,\ell}, z_{u,\ell})$ and $(\tilde{x}_{\tilde{u},\ell-1}, \tilde{z}_{\tilde{u},\ell-1})$ represent the horizontal and vertical coordinates of the $u$-th and $\tilde{u}$-th meta-atoms on the $\ell$-th and $(\ell-1)$-th layers, respectively. 
According to the Rayleigh-Sommerfeld diffraction theory, the element-wise propagation coefficient is expressed as~\cite{SIM_JSAC}
\begin{equation}
\label{eq:rayleigh-sommerfeld}
w^{\ell}_{\tilde{u},u} 
= \frac{A_e \cos \eta_{\tilde{u},u}^{\ell}}
{t_{\tilde{u},u}^{\ell}} 
\left( \frac{1}{2\pi t_{\tilde{u},u}^{\ell}} - \frac{j}{\lambda} \right) e^{j\frac{2\pi}{\lambda} t_{\tilde{u},u}^{\ell}},
\end{equation}
where $A_e$ is an area of SIM element, $\lambda$ is the wavelength, and $\eta_{\tilde{u},u}^{\ell}$ is the angle between the normal direction of the SIM layer and the propagation direction from $\tilde{u}$-th element in $(\ell-1)$-th layer to $u$-th element in $\ell$-th layer.
This formulation captures both the amplitude decay and the phase rotation due to near-field wave diffraction between layers.
Moreover, let $\phi_u^{\ell} = e^{j\theta_u^{\ell}}$ denotes the phase shift imposed by the $u$-th element in $\ell$-th layer.
We assume that the phase shift can be continuously adjusted in the interval between $0$ and $2\pi$, i.e., $\theta_u^{\ell} \in [0,2\pi)$.
Therefore, the phase configuration of the $\ell$-th layer is defined by
$\boldsymbol{\varphi}_{\ell} = [ \phi_1^{\ell}, \ldots, \phi_U^{\ell} ]^{\mathrm{T}}$,
and the corresponding diagonal phase-shift matrix is
$\boldsymbol{\Psi}_{\ell} = \mathrm{diag}(\boldsymbol{\varphi}_{\ell}) \in \mathbb{C}^{U\times U}$
where $\phi_u^{\ell}$ represents the controllable phase shift applied by the $u$-th meta-atom at $\ell$-th layer. 

Let $\mathbf{Q} = [\mathbf{q}_1, \ldots, \mathbf{q}_K] \in \mathbb{C}^{K \times U}$ represent the channel matrix from the last SIM layer to the $K$ users, where $\mathbf{q}_k \in \mathbb{C}^{1\times U}$ denotes the channel corresponding to the $k$-th user. 
Then, the end-to-end effective channel between the BS and the users via the multi-layer SIM can be written as
\begin{equation}
\mathbf{H} = 
\mathbf{Q}
\boldsymbol{\Psi}_{L}
\mathbf{W}_{L}
\boldsymbol{\Psi}_{L-1}
\cdots
\mathbf{W}_{2}
\boldsymbol{\Psi}_{1},
\end{equation}
where $\mathbf{H} = [ \mathbf{h}_1, \mathbf{h}_2, \ldots, \mathbf{h}_K ]^{\mathrm{T}}$ denotes the effective BS-SIM-UE channel matrix. 
Each user channel vector $\mathbf{h}_k$ incorporates the cumulative propagation effects of all layers, including the transmission coefficients of the SIM and the inter-layer diffraction between them.

\subsection{SIM-aided HRSMA Transmission Framework}
\label{section:sim-aided_rsma_transmission_framework}
In HRSMA, the message intended for the $k$-th user, denoted by $s_k$, is divided into three components: a common part, a group-common part, and a private part. 
The common parts of all users, denoted by $s_{\mathrm{c},k}, \forall k$, are combined and jointly encoded into a single common stream $s_{\mathrm{c}}$, which is decoded by all users. 
Moreover, the users are divided into $N$ groups, and the set of users belonging to group $n$ is denoted by $\mathcal{G}_n$. 
The group message of the $n$-th group, denoted by $s_{\mathrm{g},n}$, is constructed by combining the group-message parts of all users belonging to $\mathcal{G}_n$, and is intended to be decoded by all users within group $\mathcal{G}_n$. 
The private message of the $k$-th user is represented by $s_{\mathrm{p},k}$, which is decoded only by the corresponding user $k$. 
Accordingly, the total number of transmitted data streams is $1 {+} N {+} K$.

In contrast to conventional multiuser MISO systems that employ digital precoding at the antenna array, the proposed architecture eliminates digital spatial precoding at the transmitter. 
Each message stream is directly associated with a single transmit antenna, and no linear combination of streams is performed in the digital domain. 
After analog-to-digital conversion and power allocation, each antenna radiates its associated message stream independently. 
Therefore, without loss of generality, the number of BS antennas is set to $N_t = 1 {+} N {+} K$. 
Accordingly, the first antenna transmits the common stream, the $(1{+}n)$-th antenna transmits the $n$-th group-common stream, and the $(1{+}N{+}k)$-th antenna transmits the private stream intended for user $k$.
The vector of transmit symbols is then donated as:
\begin{equation}
    \mathbf{s}
    =
    \big[
        s_{\mathrm{c}},\,
        s_{\mathrm{g},1},\, 
        \ldots,\,
        s_{\mathrm{g},N},\,
        s_{\mathrm{p},1},\, 
        \ldots,\,
        s_{\mathrm{p},K}
    \big]^T,
\end{equation}
where all symbols are normalized such that $\mathbb{E}[|s_i|^2] = 1$. 
Moreover, the vector of transmit powers allocated to each stream is represented as:
\begin{equation}
    \mathbf{p} =
    \big[
        p_{\mathrm{c}},\,
        p_{\mathrm{g},1},\, 
        \ldots,\,
        p_{\mathrm{g},N},\,
        p_{\mathrm{p},1},\, 
        \ldots,\,
        p_{\mathrm{p},K}
    \big]^T.
\end{equation}
Therefore, the transmit signal vector radiated from the BS antenna array $\mathbf{x} = [x_{\text{c}}, x_{\text{g},1},\ldots,x_{\text{g},N},x_{\text{p},1},\ldots,x_{\text{p},K}]^T$ is then expressed as:
\begin{equation}
\mathbf{x} = \sqrt{\mathbf{p}} \odot \mathbf{s}.
\end{equation}
The total transmit power satisfies the constraint:
\begin{equation}
    p_{\text{c}} + \sum_{n=1}^{N} p_{\text{g},n} + \sum_{k=1}^{K} p_{\text{p},k} \leq P_{\mathrm{t}},
\end{equation}
where $P_{\mathrm{t}}$ is the total transmission power of the BS.

Let $\mathbf{V} = [\mathbf{v}_1,\ldots,\mathbf{v}_{N_t}] \in \mathbb{C}^{U \times N_t}$ denote the channel matrix from the BS antenna array to the first SIM layer, where $\mathbf{v}_{i} = [v_{i,1},\ldots,v_{i,U}]$ represents the channel vector from the $i$-th antenna to the first SIM layer, and the $u$-th propagation entry $v_{i,1}$ can be obtained from (\ref{eq:rayleigh-sommerfeld}). 
Hence, the received signal at user $k$ can be expressed as
\begin{align}
    y_k & = \mathbf{h}_k \mathbf{V} \mathbf{x} + n_k, \\
        & = \mathbf{h}_k \left( \mathbf{v}_1 x_{\text{c}} {+} \sum_{n=1}^{N} \mathbf{v}_{1+n} x_{\text{g},n} {+} \sum_{k=1}^{K} \mathbf{v}_{1+N+k} x_{\text{p},k} \right) {+} n_k,
\end{align}
where $n_k \sim \mathcal{CN}(0, \sigma_n^2)$ denotes the additive noise at user $k$.

Each user employs successive interference cancellation (SIC) to decode the intended streams. Specifically, user $k$ first decodes the common stream $s_{\text{c}}$, which is intended for all users, while treating all other streams as interference. After successfully decoding and removing the common message, user $k$ proceeds to decode its corresponding group-common stream $s_{\text{g},n}$ if it belongs to group $\mathcal{G}_n$, and finally decodes its private stream $s_{\text{p},k}$. Based on this decoding order, the signal-to-interference-plus-noise ratio (SINR) for the common stream at user $k$ is given by
\begin{equation}
    \gamma_{\text{c},k} = 
    \frac{ p_c |\mathbf{h}_k \mathbf{v}_1|^2 }
    { \sum_{n=1}^{N} p_{\text{g},n} |\mathbf{h}_k \mathbf{v}_{1{+}n}|^2 {+} \sum_{k=1}^{K} p_{\text{p},k} |\mathbf{h}_k \mathbf{v}_{1{+}N{+}k}|^2 {+} \sigma_n^2 }.
\end{equation}
For the group-common stream associated with group $n$, the SINR at user $k \in \mathcal{G}_n$ is expressed as
\begin{equation}
    \gamma_{\mathrm{g},n,k} = 
    \frac{ p_{\mathrm{g},n} |\mathbf{h}_k \mathbf{v}_{1{+}n}|^2 }
    { \sum_{\substack{l \neq n}}^{N} p_{\mathrm{g},l} |\mathbf{h}_k \mathbf{v}_{1{+}l}|^2 
    {+} \sum_{k=1}^{K} p_{\mathrm{p},k} |\mathbf{h}_k \mathbf{v}_{1{+}N{+}k}|^2 
    {+} \sigma_n^2 }.
\end{equation}
Finally, the SINR for the private stream of user $k$ is given by
\begin{equation}
    \gamma_{\mathrm{p},k} = 
    \frac{ p_{\mathrm{p},k} |\mathbf{h}_k \mathbf{v}_{1{+}N{+}k}|^2 }
    { \sum_{\substack{l \neq k}}^{K} p_{\mathrm{p},l} |\mathbf{h}_k \mathbf{v}_{1{+}N{+}l}|^2 
    {+} \sigma_n^2 }.
\end{equation}

Based on the previously described decoding order, the achievable rate of each stream is determined according to the corresponding SINR. 
Since the common stream $s_{\mathrm{c}}$ must be successfully decoded by all users, its rate is limited by the user who experiences the lowest SINR. 
Hence, the achievable rate for the common stream is expressed as
\begin{equation}
    R_{\mathrm{c}} = \min_{k} \log_2 \!\left( 1 + \gamma_{\mathrm{c},k} \right).
\end{equation}
Similarly, for the group-common stream $s_{\mathrm{g},n}$, which is intended for a specific subset of users $\mathcal{G}_n$, 
the achievable rate is restricted by the user with the lowest SINR within that group. 
Accordingly, it can be written as
\begin{equation}
    R_{\mathrm{g},n} = 
    \min_{k \in \mathcal{G}_n} 
    \log_2 \!\left( 1 + \gamma_{\mathrm{g},n,k} \right).
\end{equation}
For the private stream $s_{\mathrm{p},k}$, which is decoded only by user $k$, 
the achievable rate is directly determined by its corresponding SINR, given by
\begin{equation}
    R_{\mathrm{p},k} = \log_2 \!\left( 1 + \gamma_{\mathrm{p},k} \right).
\end{equation}

Finally, the total achievable rate of user $k$ is obtained by aggregating the contributions from the common, group-common, and private streams. 
We assume that the common rate $R_{\mathrm{c}}$ is equally divided among all users, 
whereas each group-common rate $R_{\mathrm{g},n}$ is equitably shared among the users in group $\mathcal{G}_n$. 
Thus, the overall achievable rate of user $k$ is expressed as
\begin{equation}
    R_k = 
    \frac{R_{\mathrm{c}}}{K} 
    + 
    \sum_{n=1}^{N} 
    \mathbb{I}[k \in \mathcal{G}_n] 
    \frac{R_{\mathrm{g},n}}{|\mathcal{G}_n|} 
    + 
    R_{\mathrm{p},k},
\end{equation}
where $\mathbb{I}[\cdot]$ denotes the indicator function, which equals $1$ if user $k$ belongs to group $\mathcal{G}_n$ and $0$ otherwise. 

\section{Problem Formulation}
\label{section:problem_formulation}
In this section, we formulate an optimization problem that aims to maximize the minimum user rate by jointly optimizing the SIM phase shifts, the user grouping structure, and the power allocation among the message streams.
Specifically, the optimization variables include the set of layer-wise phase-shift vectors 
$\{\boldsymbol{\varphi}_\ell\}_{\ell=1}^{L}$, the transmit power vector $\mathbf{p}$, and the user grouping sets $\{\mathcal{G}_n\}_{n=1}^{N}$.
The optimization problem can be formulated as
\begin{subequations}
\label{eq:opt_problem}
\begin{align}
    \max_{\{\boldsymbol{\varphi}_\ell\},\, \mathbf{p},\, \{\mathcal{G}_n\}} 
    & \quad \min_{k} R_k  \\
    \text{s.t.} 
    & \quad |\phi_u^{\ell}| = 1, \forall u, \ell. \label{eq:constraint_phase}\\
    & \quad p_{\text{c}} + \sum_{n=1}^{N} p_{\text{g},n} + \sum_{k=1}^{K} p_{\text{p},k} \leq P_{\mathrm{t}}, \label{eq:constraint_power_ub}\\
    & \quad p_{\text{c}} \ge 0 , p_{\text{g},n} \ge 0, p_{\text{p},k} \ge 0, \forall n, k, \label{eq:constraint_power_nonzero}\\
    & \quad \bigcup_{n=1}^{N} \mathcal{G}_n, \quad \mathcal{G}_i \cap \mathcal{G}_j = \emptyset, \; \forall i \ne j.\label{eq:constraint_group}
\end{align}
\end{subequations}
The constraint (\ref{eq:constraint_phase}) enforces the unit-modulus property of each programmable meta-atom in the $\ell$-th SIM layer.  
The constraint (\ref{eq:constraint_power_ub}) ensures that the total transmit power does not exceed the total transmission power.
The constraint (\ref{eq:constraint_power_nonzero}) guarantees that all power allocation variables are non-negative. 
Constraint (\ref{eq:constraint_group}) ensures that the user sets $\{\mathcal{G}_n\}$ form a valid partition of the entire user set $\{1,2,\ldots,K\}$, i.e., each user belongs to exactly one group and no two groups share the same user. 

This optimization problem is non-convex owing to the coupling between the SIM phase shift, the transmit power vector, and the user grouping structure. 
Moreover, the unit-modulus constraints in the SIM phase shift and discrete user grouping variables further increase the combinatorial complexity of the problem. 
To address this, an alternating optimization-based algorithm is developed in the next section, which efficiently decomposes the joint optimization into several tractable subproblems.

\section{Joint Design of SIM Phase Shift, Power Allocation, and User Grouping}
\label{section:joint_optimization}
To solve the non-convex problem formulated in Section~\ref{eq:opt_problem}, we develop an alternating optimization (AO) framework that iteratively updates the SIM phase-shift matrices, the power allocation vector, and the user grouping structure. 
The algorithm aims to maximize the minimum achievable rate among all users while satisfying the total power, user grouping, and unit-modulus constraints.

\begin{algorithm}[t]
\caption{Alternating Optimization for SIM-aided HRSMA}
\label{alg:AO_framework}
\begin{algorithmic}[1]
\STATE \textbf{Input:} Channel matrices $\{\mathbf{H}, \mathbf{V}\}$, user and group counts $(K,N)$
\STATE Randomly initialize SIM phase-shift matrices $\{\boldsymbol{\varphi}_\ell^{(0)}\}$
\STATE Randomly initialize transmit power vector $\mathbf{p}^{(0)}$
\STATE Form $\mathbf{H}_{\!\mathrm{eff}}$ and extract user features $\{\tilde{\mathbf{f}}_k\}_{k=1}^K$
\STATE Apply $k$-means clustering to $\{\tilde{\mathbf{f}}_k\}$ 
to obtain the initial user grouping $\{\mathcal{G}_n^{(0)}\}_{n=1}^N$

\FOR{$t = 1$ to $I_{\max}$}
    \STATE Optimize $\{\boldsymbol{\varphi}_\ell^{(t)}\}$ via \textbf{Algorithm~\ref{alg:phase_spsa}}
    \STATE Optimize $\mathbf{p}^{(t)}$ via \textbf{Algorithm~\ref{alg:power_spsa}}
    \STATE Update grouping $\{\mathcal{G}_n^{(t)}\}$ via \textbf{Algorithm~\ref{alg:group_opt}}
    \STATE Evaluate minimum rate:  \\ $R_{\min}^{(t)} = \min_k R_k(\mathbf{p}^{(t)}, \{\boldsymbol{\varphi}_\ell^{(t)}\}, \{\mathcal{G}_n^{(t)}\})$
    \IF{$|R_{\min}^{(t)} - R_{\min}^{(t-1)}| < \epsilon$}
        \STATE \textbf{break}
    \ENDIF
\ENDFOR
\STATE \textbf{Output:} $\{\boldsymbol{\varphi}_\ell^{(t)}\}$, $\mathbf{p}^{(t)}$, $\{\mathcal{G}_n^{(t)}\}$, and $R_{\min}^{(t)}$
\end{algorithmic}
\end{algorithm}

\subsection{Algorithm Overview}
The proposed AO algorithm initializes all variables and then iteratively refines them. 
In each iteration, the algorithm alternately optimizes the following three components:
\begin{enumerate}
    \item SIM phase optimization:  
    The phase-shift vectors $\{\boldsymbol{\varphi}_\ell\}$ are optimized using a simultaneous perturbation stochastic approximation (SPSA)~\cite{SPSA} 
    to maximize the minimum user rate.
    \item Power allocation optimization:  
    The transmit power vector $\mathbf{p}$ is optimized under the total power constraint 
    using an SPSA-based stochastic ascent scheme.
    \item User grouping optimization:  
    The user groups $\{\mathcal{G}_n\}$ are updated using a fast clustering-based heuristic 
    followed by local greedy refinement to improve the bottleneck user rate.
\end{enumerate}
This procedure repeats until the minimum user rate converges or the maximum iteration count $I_{\max}$ is reached.
This algorithm is presented in Algorithm~\ref{alg:AO_framework}.

\begin{algorithm}[t]
\caption{SPSA-based SIM Phase Shift Optimization}
\label{alg:phase_spsa}
\begin{algorithmic}[1]
\STATE \textbf{Input:} Number of iterations $t$ in \textbf{Algorithm~\ref{alg:AO_framework}},\\
$\mathbf{p}^{(t)}$, $\{\mathcal{G}_n^{(t)}\}$, and $\boldsymbol{\varphi}^{(t)}$ in $t$-th iteration
\STATE Generate perturbation $\boldsymbol{\Delta}^{(t)} \in \{\pm 1\}^{LU}$
\STATE Form perturbed vectors:
$\boldsymbol{\varphi}^{\pm} = \mathcal{P}\!\big(\boldsymbol{\varphi}^{(t)} \pm c_{\text{p},t}\boldsymbol{\Delta}^{(t)}\big)$
\STATE Evaluate objective values: 
$f_\text{p}^{+} = f_\text{p}(\boldsymbol{\varphi}^{+}), \quad f_\text{p}^{-} = f_\text{p}(\boldsymbol{\varphi}^{-})$
\STATE Compute gradient estimate: $\widehat{\nabla} f_\text{p}(\boldsymbol{\varphi}^{(t)}) = 
\frac{f_\text{p}^{+}-f_\text{p}^{-}}{2c_{\text{p},t}} \,\boldsymbol{\Delta}^{(t)\,-1}$
\STATE Update phase vector with projection:\\
$\boldsymbol{\varphi}^{(t+1)} = 
\mathcal{P}\!\big(\boldsymbol{\varphi}^{(t)} + a_{\text{p},t}\widehat{\nabla} f_\text{p}(\boldsymbol{\varphi}^{(t)})\big)$
\STATE \textbf{Output:} $\boldsymbol{\varphi}^{(t)}$
\end{algorithmic}
\end{algorithm}

\subsection{SIM Phase Shift Optimization}
For a fixed power allocation and user grouping, the SIM phase-shift matrices $\{\boldsymbol{\varphi}_\ell\}_{\ell=1}^{L}$ are optimized to maximize the minimum achievable user rate:
\begin{align}
    \max_{\{\boldsymbol{\varphi}_\ell\}} \quad & f_\text{p}(\{\boldsymbol{\varphi}_\ell\}) \\
    \quad \text{s.t.} \quad & f_\text{p}(\{\boldsymbol{\varphi}_\ell\}) = \min_{k} R_k, \\
                            & |\phi_u^{\ell}| = 1, \; \forall u, \ell.
    \label{eq:spsa_phase_opt}
\end{align}
Given that the objective is a high-dimensional, non-convex function, SPSA~\cite{SPSA} is adopted to efficiently estimate gradients via two random perturbations per iteration.
The objective function of the optimization problem (\ref{eq:opt_problem}) is not smooth owing to the inclusion of $\min_k$, and therefore, its explicit analytical differentiation cannot be computed. 
SPSA enables the update of the optimization variable values by numerically approximating the gradient.
Moreover, a standard finite-difference gradient would need $2LU$ objective calls per iteration to derive the gradient. SPSA needs only two, which is critical when the number of elements and layers is large.
Let $\boldsymbol{\varphi} = [\boldsymbol{\varphi}_1,\ldots,\boldsymbol{\varphi}_L]$ stack all phase shift variables. 
At iteration $t$, a random perturbation vector $\boldsymbol{\Delta}^{(t)} \in \{\pm 1\}^{LU}$ is generated, where each element is drawn independently from a Rademacher distribution. Two perturbed angle vectors are then formed as
\begin{equation}
    \boldsymbol{\varphi}^{\pm} = \mathcal{P}\!\big(\boldsymbol{\varphi}^{(t)} \pm c_{\text{p},t} \boldsymbol{\Delta}^{(t)}\big),
\end{equation}
where $\mathcal{P}(\cdot)$ denotes element-wise wrapping onto $[0,2\pi)$ to enforce the unit-modulus constraint. The objective function is evaluated at these two points to obtain $f_\text{p}^{+} = f_\text{p}(\boldsymbol{\varphi}^{+})$ and $f_\text{p}^{-} = f_\text{p}(\boldsymbol{\varphi}^{-})$.
The two-sided stochastic gradient estimate is then given by
\begin{equation}
    \widehat{\nabla} f_\text{p}(\boldsymbol{\varphi}^{(t)}) 
    = \frac{f_\text{p}^{+} - f_\text{p}^{-}}{2c_{\text{p},t}}\,\boldsymbol{\Delta}^{(t)\,-1},
\end{equation}
where $\boldsymbol{\Delta}^{(t)\,-1}$ denotes the element-wise inverse of $\boldsymbol{\Delta}^{(t)}$.
Subsequently, the phase vector is updated along the estimated ascent direction according to
\begin{equation}
    \boldsymbol{\varphi}^{(t+1)} 
    = \mathcal{P}\!\left(\boldsymbol{\varphi}^{(t)} + a_{\text{p},t}\,\widehat{\nabla} f_\text{p}(\boldsymbol{\varphi}^{(t)})\right),
\end{equation}
where $a_{\text{p},t}$ and $c_{\text{p},t}$ are the step-size and perturbation sequences for SIM phase shift optimization, respectively.
This algorithm is presented in Algorithm~\ref{alg:phase_spsa}.

\begin{algorithm}[t]
\caption{SPSA-based Power Allocation Optimization}
\label{alg:power_spsa}
\begin{algorithmic}[1]
\STATE \textbf{Input:} Number of iterations $t$ in \textbf{Algorithm~\ref{alg:AO_framework}},\\
$\mathbf{p}^{(t)}$, $\{\mathcal{G}_n^{(t)}\}$, and $\boldsymbol{\varphi}^{(t)}$ in $t$-th iteration
\STATE Generate perturbation $\boldsymbol{\Delta}^{(t)} \in \{\pm 1\}^{1+N+K}$
\STATE Form perturbed power vectors:
$\mathbf{p}^{\pm} = \Pi\!\big(\mathbf{p}^{(t)} \pm c_{\text{a},t}\boldsymbol{\Delta}^{(t)}\big)$
\STATE Evaluate objective values:
$f_\text{a}^{+} = f_\text{a}(\mathbf{p}^{+}), \quad f_\text{a}^{-} = f_\text{a}(\mathbf{p}^{-})$
\STATE Compute gradient estimate:
$\widehat{\nabla} f_\text{a}(\mathbf{p}^{(t)}) = 
\frac{f_\text{a}^{+}-f_\text{a}^{-}}{2c_{\text{a},t}} \,\boldsymbol{\Delta}^{(t)\,-1}$
\STATE Update power allocation with projection:\\
$\mathbf{p}^{(t+1)} = 
\Pi\!\Big(\mathbf{p}^{(t)} + a_{\text{a},t}\widehat{\nabla} f_\text{a}(\mathbf{p}^{(t)})\Big)$
\STATE where
$\Pi(\mathbf{z}) = 
\frac{P_{\mathrm{t}}}{\max(\mathbf{1}^{\top}\max(\mathbf{z},0),\,\varepsilon)}\,\max(\mathbf{z},0)$
\STATE \textbf{Output:} $\mathbf{p}^{(t)}$
\end{algorithmic}
\end{algorithm}

\subsection{Power Allocation Optimization}
\label{subsec:power_spsa}
Given the SIM configuration $\{\boldsymbol{\varphi}_\ell\}$ and the user grouping $\{\mathcal{G}_n\}$,
we optimize the power allocation among the $1{+}N{+}K$ logical streams to maximize the minimum user rate.
\begin{align}
    \max_{\mathbf{p}} \quad & f_\text{a}(\mathbf{p}) \\
    \quad \text{s.t.} \quad & f_\text{a}(\mathbf{p}) = \min_{k} R_k\\
                            & \quad p_{\text{c}} + \sum_{n=1}^{N} p_{\text{g},n} + \sum_{k=1}^{K} p_{\text{p},k} \leq P_{\mathrm{t}},\\
                            & \quad p_{\text{c}} \ge 0 , p_{\text{g},n} \ge 0, p_{\text{p},k} \ge 0, \forall n, k.
    \label{eq:power_opt_spsa}
\end{align}
Similar to the phase step, SPSA estimates the gradient from the two function evaluations as follows:
\begin{equation}
    \mathbf{p}^{\pm}=\Pi\!\big(\mathbf{p}^{(t)} \pm c_{\text{a},t}\boldsymbol{\Delta}^{(t)}\big),
\end{equation}
where $\Pi(\cdot)$ is an operation that normalizes the power allocation so that the total transmit power does not exceed the constraint, and $\varepsilon$ avoids division by zero, defined as:
\begin{equation}
    \Pi(\mathbf{z})=\frac{P_{\mathrm{t}}}{\max(\mathbf{1}^{\top}\max(\mathbf{z},0),\,\varepsilon)}\,\max(\mathbf{z},0).
\end{equation}
Evaluate $f_\text{a}^{\pm}=f_\text{a}(\mathbf{p}^{\pm})$ and build the two-sided SPSA gradient estimate
\begin{equation}
\widehat{\nabla} f_\text{a}(\mathbf{p}^{(t)})=\frac{f_\text{a}^{+}-f_\text{a}^{-}}{2c_{\text{a},t}}\;\boldsymbol{\Delta}^{(t)\,-1},
\end{equation}
followed by the projected ascent step
\begin{equation}
\mathbf{p}^{(t+1)}=\Pi\!\Big(\mathbf{p}^{(t)} + a_{\text{a},t}\,\widehat{\nabla} f_\text{a}(\mathbf{p}^{(t)})\Big),
\label{eq:spsa_power_update}
\end{equation}
where $a_{\text{a},t}$ and $c_{\text{a},t}$ are the step-size and perturbation sequences for power allocation optimization, respectively.
This algorithm is presented in Algorithm~\ref{alg:power_spsa}.

\begin{algorithm}[t]
\caption{User Grouping via $k$-means Initialization and Local Greedy Refinement}
\label{alg:group_opt}
\begin{algorithmic}[1]
\STATE \textbf{Input:} Number of iterations $t$ in \textbf{Algorithm~\ref{alg:AO_framework}},\\
$\mathbf{p}^{(t)}$, $\{\mathcal{G}_n^{(t)}\}$, and $\boldsymbol{\varphi}^{(t)}$ in $t$-th iteration

\FOR{$i = 1$ to $I_{\mathrm{ref}}$}
    \STATE Identify bottleneck user $k^{\star} \in \arg\min_{k} R_k$
    \STATE Let $g_0$ be the current group index with $k^{\star} \in \mathcal{G}_{g_0}$
    \STATE \text{flag} $\leftarrow$ \textbf{true}, ${R}_{\min} \leftarrow \min_{k} R_k$
    \FOR{each $g \in \{1,\ldots,N\}\setminus\{g_0\}$}
        \STATE Construct candidate partition $\{\widetilde{\mathcal{G}}_n\}$: \\
        $\widetilde{\mathcal{G}}_{g_0} = \mathcal{G}_{g_0} \setminus \{k^{\star}\}$ \\
        $\widetilde{\mathcal{G}}_{g}   = \mathcal{G}_{g} \cup \{k^{\star}\}$ \\
        $\widetilde{\mathcal{G}}_{n}   = \mathcal{G}_{n} \ \ (n\notin\{g_0,g\})$ \\
        \STATE Evaluate candidate minimum rate: \\
        $\widetilde{R}_{\min} = \min_{k} R_k\big(\{\widetilde{\mathcal{G}}_n\}\big)$
        \IF{$\widetilde{R}_{\min} > {R}_{\min}$}
            \STATE ${R}_{\min} \leftarrow \widetilde{R}_{\min}$, 
            $\{{\mathcal{G}}_n\}\leftarrow\{\widetilde{\mathcal{G}}_n\}$,
            $\text{flag}\leftarrow\textbf{true}$
        \ENDIF
    \ENDFOR
    \IF{$\text{flag}=\textbf{false}$}
        \STATE \textbf{break} 
    \ENDIF
\ENDFOR
\STATE \textbf{Output:} $\{\mathcal{G}_n\}$
\end{algorithmic}
\end{algorithm}

\subsection{User Grouping Optimization}
\label{subsec:group_update}
Given $\mathbf{p}$ and ${\boldsymbol{\Psi}_\ell}$, the user grouping $\mathcal{G}_n$ is updated to improve the minimum user rate.
Exhaustive enumeration of all possible partitions is infeasible; hence, we adopt a two-step heuristic.
Before starting the AO iterations, $k$-means clustering is performed to obtain an effective initialization of the user groups.
During each AO iteration, a local greedy refinement is applied, which directly targets the bottleneck user to further enhance the minimum rate.

\subsubsection{k-Means initialization}
The group-common rate $R_{\mathrm{g},n}$ is determined by the weakest user within the group. 
Through phase control by the SIM, spatial separation of inter-group and intra-group messages is achieved by exploiting channel orthogonality. 
Therefore, when users within the same group exhibit similar channel characteristics, their received power and interference structure for the group-common stream tend to be similar, which prevents the minimum rate from dropping drastically. 
To obtain an initial grouping that mitigates such degradation, $k$-means clustering is applied so that users with similar channel features are assigned to the same cluster.

Let $\mathbf{H}_{\!\mathrm{eff}} = \mathbf{HV}\in \mathbb{C}^{K\times(1{+}N{+}K)}$ be the effective channel between users and BS antennas. 
For user $k$, define a feature vector by concatenating the real and imaginary parts (or their magnitudes) of the $k$-th row:
\begin{equation}
\mathbf{z}_k = \mathbf{H}_{\!\mathrm{eff}}(k,:),\quad
\mathbf{f}_k = 
\begin{bmatrix}
\mathrm{Re}\{\mathbf{z}_k\} \\
\mathrm{Im}\{\mathbf{z}_k\}
\end{bmatrix}
\;\;\text{or}\;\;
\mathbf{f}_k \;=\; |\mathbf{z}_k| .
\end{equation}
Optionally, normalize $\tilde{\mathbf{f}}_k = \mathbf{f}_k / |\mathbf{f}_k\|_2$. We then cluster $\{\tilde{\mathbf{f}}_k\}_{k=1}^{K}$ into $N$ clusters via $k$-means (with multiple replicates), yielding the initial partition $\{\mathcal{G}_n^{(0)}\}_{n=1}^{N}$. 

\subsubsection{Local Greedy Refinement}
Starting from $\{\mathcal{G}_n^{(0)}\}$, we iteratively improve the minimum rate by reassigning the current bottleneck user. In $t$-th iteration, let the bottleneck user be $k^{\star} \in \arg\min_{k} R_k$ and the user $k^{\star}$ belongs to the group $g_0$, i.e., $k^{\star} \in \mathcal{G}_{g_0}$.
For each $g \in \{1,\ldots,N\}\setminus\{g_0\}$, construct a candidate partition $\{\widetilde{\mathcal{G}}_n\}$ by moving $k^{\star}$ from $g_0$ to $g$:
\begin{align}
\widetilde{\mathcal{G}}_{g_0} &= \mathcal{G}_{g_0}\setminus\{k^{\star}\}, \\
\widetilde{\mathcal{G}}_{g}   &= \mathcal{G}_{g} \cup \{k^{\star}\}, \\
\widetilde{\mathcal{G}}_{n}   &= \mathcal{G}_{n}\; (n\notin\{g_0,g\}).
\end{align}
Then, the minimum rate for the candidate partition, denoted by $\tilde{R}_{\min}$, is evaluated. 
If $\tilde{R}_{\min} > R_{\min}$, where $R_{\min}$ denotes the current minimum rate in the iteration, the move is accepted and both the current partition and $R_{\min}$ are updated accordingly. 
This procedure is repeated for at most $I_{\mathrm{ref}}$ refinement rounds, 
or terminated earlier if no improving move is found. 
Finally, the grouping result and the corresponding rate obtained after the refinement 
are adopted as the output of the user grouping optimization step 
in the $t$-th iteration of the alternating optimization algorithm.
This algorithm is presented in Algorithm~\ref{alg:group_opt}.

\begin{figure}[t]
    \centering
    \includegraphics[width=0.95\hsize]{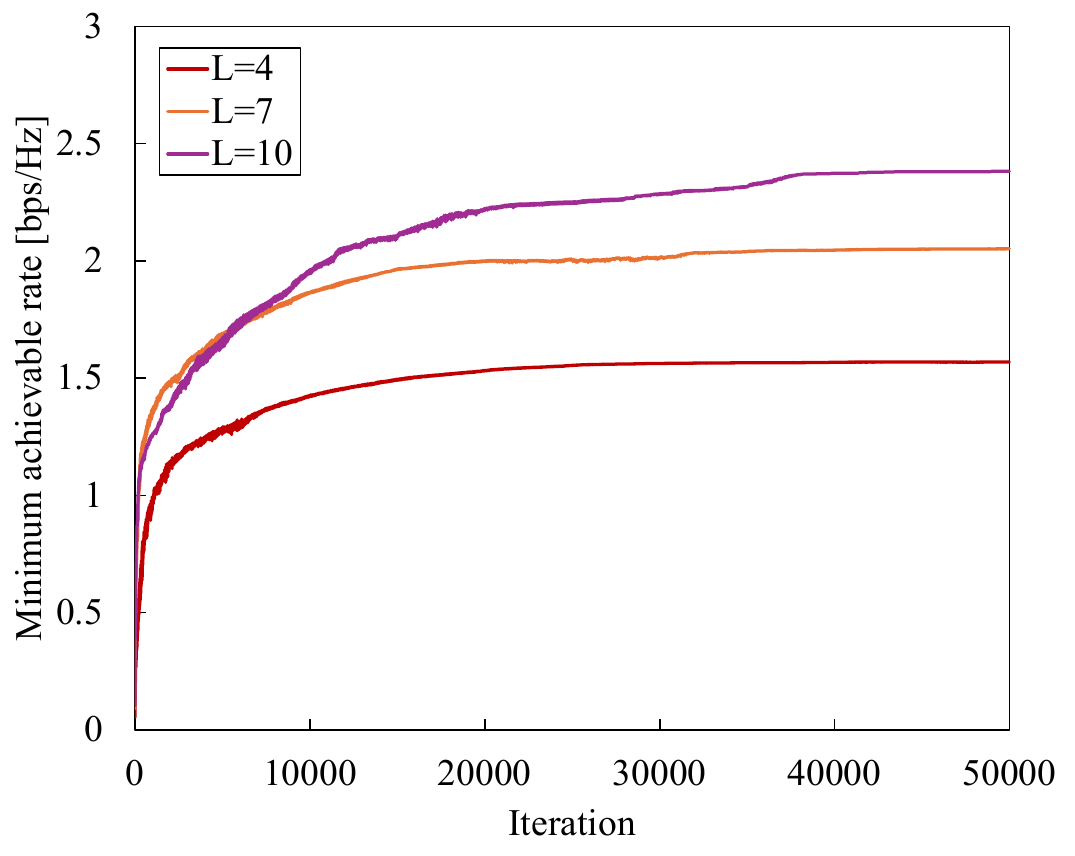}
    \caption{Convergence behavior of Algorithm~\ref{alg:AO_framework}.}
    \label{fig:convergence}
\end{figure}

\subsection{Convergence and Complexity}
Each AO iteration monotonically improves or maintains the best-so-far minimum rate. 
Both SPSA substeps involve only two function evaluations per iteration, and each update guarantees feasibility through projection operations.
The grouping refinement always increases $R_{\min}$ until a local optimum is reached. 
The overall per-iteration complexity scales as $\mathcal{O}(L U^2 + (1+N+K) + KN)$ dominated by effective-channel computation and rate evaluation. 
Hence, the proposed AO framework achieves a favorable balance between rate performance and computational efficiency.

Fig.~\ref{fig:convergence} illustrates the convergence behavior of the Algorithm~\ref{alg:AO_framework}, 
where the minimum achievable rate is plotted against the iteration index. 
Results are shown for three different numbers of the SIM layers, $L \in \{4,7,10\}$. 
In all cases, the objective function (i.e., the minimum user rate) monotonically increases with each iteration, 
demonstrating the stability and effectiveness of the proposed joint optimization scheme.
At the early stage ($<5000$ iterations), the minimum rate improves sharply, 
as the SPSA-based phase optimization and power allocation steps rapidly identify better local directions of ascent. 
Subsequently, the improvement gradually tapers and the algorithm converges smoothly to a steady-state value. 
This monotonic convergence behavior verifies that each update of the SIM phase shifts, power allocation, and grouping does not degrade the overall objective function.
Moreover, a larger number of SIM layers $L$ leads to a higher converged minimum rate. 
This is because additional layers increase the controllable degrees of freedom in wavefront manipulation, 
allowing finer adjustment of the effective channel $\mathbf{H}_{\mathrm{eff}}$ to balance the received power among users. 
Nevertheless, the convergence speed slightly decreases with increasing $L$, 
as the dimensionality of the phase-shift space grows and the optimization landscape becomes more complex.
Overall, the results confirm that the proposed AO algorithm converges reliably within a reasonable number of iterations and consistently enhances the minimum user rate, validating the robustness of the SPSA-based optimization framework.

\begin{figure}[t]
	\centering
	\includegraphics[width=1.0\hsize]{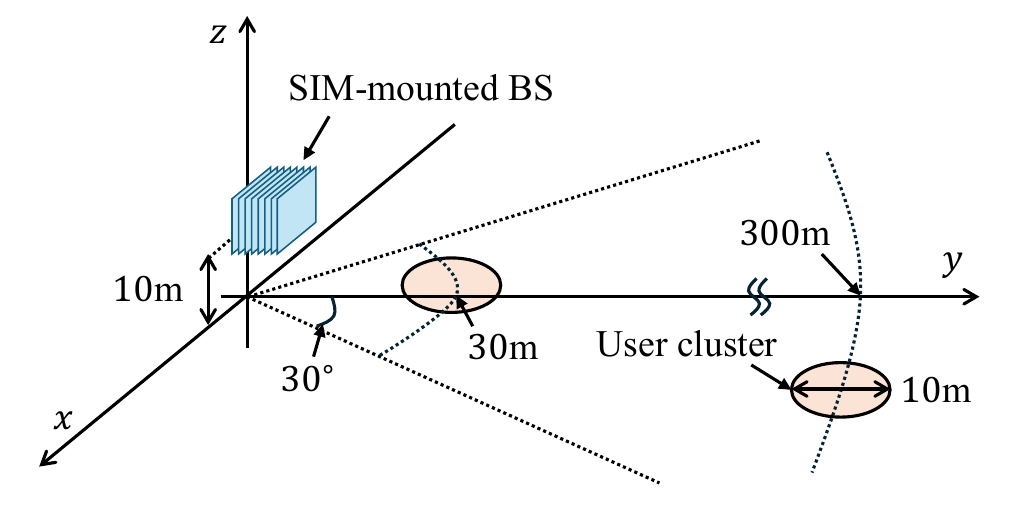}
	\caption{Simulation environment.}
	\label{fig:simulation_environment}
\end{figure}

\begin{table}[t]
\centering
\caption{Simulation Parameters}
\label{table:simulation_parameter}
\begin{tabular}{l | l}

\textbf{Parameter} & \textbf{Value} \\
\hline \hline
BS location & $(0, 0, 10)$~m \\
Transmission power & 20~dBm \\
Noise power & $-94$~dBm \\
Number of users ($K$) & 6 \\
Number of groups ($N$) & 2 \\
Number of SIM layers ($L$) & 7 \\
Number of SIM elements per layer ($U$) & 64 \\
Carrier frequency & 28~GHz \\
Element spacing (BS/SIM) & $\lambda/2$ \\
SIM layer spacing ($d_\ell$) & $\lambda/4$ \\
Height of users & 1.5~m \\
Rician factor for SIM--UE channel $K_\text{R}$& 13~dB \\
\end{tabular}
\end{table}

\section{Performance Evaluation}
\label{section:performance_evaluation}
\subsection{Simulation Scenario}
In this section, we evaluate the effectiveness of the proposed SIM-aided HRSMA architecture. 
We consider a single-cell downlink network, as illustrated in Fig.~\ref{fig:simulation_environment}, 
where a SIM is embedded inside the radome of the base station (BS). 
The BS antenna is located at the center point $(0, 0, 10)$~m and covers a $60^{\circ}$ sector as shown in Fig.~\ref{fig:simulation_environment}. 
It serves $K$ single-antenna users that are randomly distributed within this sector, 
forming two clusters positioned on circular arcs with radii of $30$~m and $300$~m from the BS, respectively. 
Each cluster occupies a $10$~m-diameter region whose center is along the corresponding arc. 
This clustered distribution is designed to verify the ability of the hierarchical rate-splitting mechanism to improve the performance of bottleneck users suffering from high path loss. 
Unless otherwise stated, the simulation parameters are summarized in Table~\ref{table:simulation_parameter}.

We assume that the SIM--UE channel matrix $\mathbf{Q}$ follows Rician fading. 
Both the BS antenna array and the SIM metasurface are configured as uniform planar arrays (UPAs) 
with half-wavelength spacing between adjacent elements. 
Accordingly, the channel between the SIM and the $k$-th user is expressed as
\begin{equation}
\mathbf{q}_k
= 
\Lambda_k
\left(
\sqrt{\frac{K_{\mathrm{R}}}{K_{\mathrm{R}} + 1}} \,
\mathbf{a}_{U}(\xi_k,\zeta_k)
+ 
\sqrt{\frac{1}{K_{\mathrm{R}} + 1}} \,
\mathbf{q}_k^{\mathrm{NLoS}}
\right),
\label{eq:qk_Rician}
\end{equation}
where $K_{\mathrm{R}}$ denotes the Rician factor, and $\Lambda_k$ represents the large-scale path loss between the SIM and user~$k$. 
The vector $\mathbf{a}_{U}(\xi_k,\zeta_k)$ denotes the array steering vector of a $U_a \!\times\! U_a$-element uniform planar array (UPA), 
with $\xi_k$ and $\zeta_k$ representing the elevation and azimuth angles, respectively, between the SIM and the $k$-th user. 
The NLoS component $\mathbf{q}_k^{\mathrm{NLoS}}$ models the scattered paths, whose entries are independently drawn from 
a circularly symmetric complex Gaussian distribution, i.e., $\mathcal{CN}(0,1)$.
The BS antenna array is configured as a UPA with 
$M_{\mathrm{BS}} = \lceil \sqrt{N_t} \, \rceil$ elements per dimension, 
resulting in a total of $N_{\mathrm{BS}} = M_{\mathrm{BS}}^2$ antennas satisfying $N_{\mathrm{BS}} \ge N_t$. 
Only $N_t$ antennas corresponding to the active message streams are activated during transmission.
The simulation parameters are summarized in Table~\ref{table:simulation_parameter}, and they are adopted throughout the simulations unless stated otherwise.

We compare the performance of the proposed method with the following six benchmark schemes:
\begin{itemize}
    \item {SIM-aided HRSMA (proposed):}  
    This is the proposed hierarchical rate-splitting architecture assisted by a SIM.  
    The number of BS antennas is set to $N_t = 1 + N + K$, where each antenna is directly associated with one message stream.  
    No digital precoding is performed at the BS.  
    The SIM acts as an analog precoder in the wave domain, aiming to spatially diagonalize the effective channel of each stream.  
    The transmit power allocation, SIM phase shifts, and user grouping are jointly optimized by Algorithm~\ref{alg:AO_framework}.
    \item {SIM-aided RSMA (non-hierarchical):}  
    This is a simplified version of the proposed system with no grouping, i.e., $N = 0$. 
    The number of BS antennas is $N_t = 1 + K$, and the optimization follows Algorithm~\ref{alg:AO_framework} 
    while skipping the grouping step.
    \item {Hybrid Beamforming-based HRSMA:}  
    The conventional HRSMA system implemented using the hybrid beamforming (HBF) architecture 
    shown in Fig.~\ref{fig:SIMandHBF_HRSMA}(a).  
    Both the digital and analog precoding vectors, as well as per-stream power allocations, are jointly optimized.  
    Unless otherwise specified, the number of BS antennas is set to $N_t = 1 + N + K$ for fair comparison.
    \item {Hybrid Beamforming-based RSMA:}  
    The RSMA system uses the same hybrid beamforming architecture as the preceding system, with $N_t = 1 + K$. 
    The digital and analog precoding vectors and power allocations are optimized accordingly.
    \item {Non-precoding HRSMA:}  
    The BS has $N_t = 1 + N + K$ antennas, each assigned to one message stream without any precoding.  
    The signals are transmitted after power allocation, corresponding to the hierarchical message structure of HRSMA.
    \item {Non-precoding RSMA:}  
    Similar to the preceding but corresponding to a non-hierarchical RSMA system with $N_t = 1 + K$.  
    Each antenna transmits one message stream independently after power allocation.
\end{itemize}
The simulations are conducted over $1800$ different positions of clusters and users, and the results represent their average values.

\subsection{Results}

\begin{figure}[t]
    \centering
    \includegraphics[width=0.95\hsize]{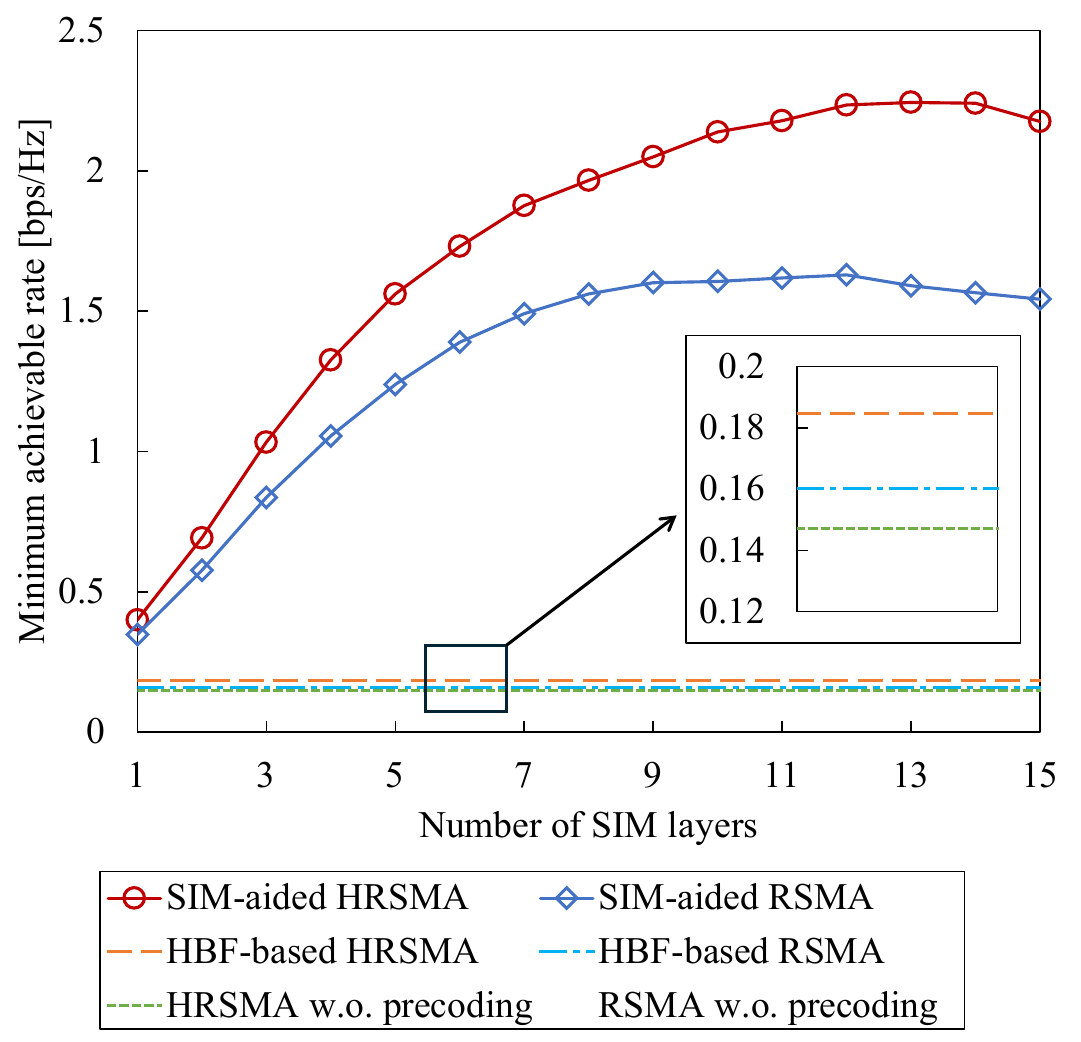}
    \caption{Minimum achievable rate versus the number of SIM layers.}
    \label{fig:num_layer}
\end{figure}

Fig.~\ref{fig:num_layer} illustrates the minimum achievable rate as a function of the number of SIM layers $L$. 
Both SIM-HRSMA and SIM-RSMA show significant improvement with increasing $L$.
This demonstrates that the multi-layer SIM structure effectively enhances wave manipulation diversity and spatial degrees of freedom. 
Moreover, as the number of SIM layers increases, the performance gain eventually saturates.
This is because once the accumulated phase diversity reaches a sufficient level to span the dominant spatial modes of the propagation environment, further layers become redundant and do not provide additional degrees of freedom.
In contrast, the conventional RSMA/HRSMA systems with or without digital precoding achieve only marginal rates, highlighting the substantial gain introduced by the SIM.
Notably, SIM-HRSMA consistently outperforms SIM-RSMA owing to the additional inter-group rate-splitting layer, which provides improved fairness among users.
The enhanced spatial separability introduced by the SIM facilitates more distinct group formation, allowing users with similar channel characteristics to be more effectively clustered. 
This alignment between physical-layer wave control and user grouping yields a better balance of intra-group coherence and inter-group isolation, resulting in higher minimum user rates.

\begin{figure}[t]
    \centering
    \includegraphics[width=0.95\hsize]{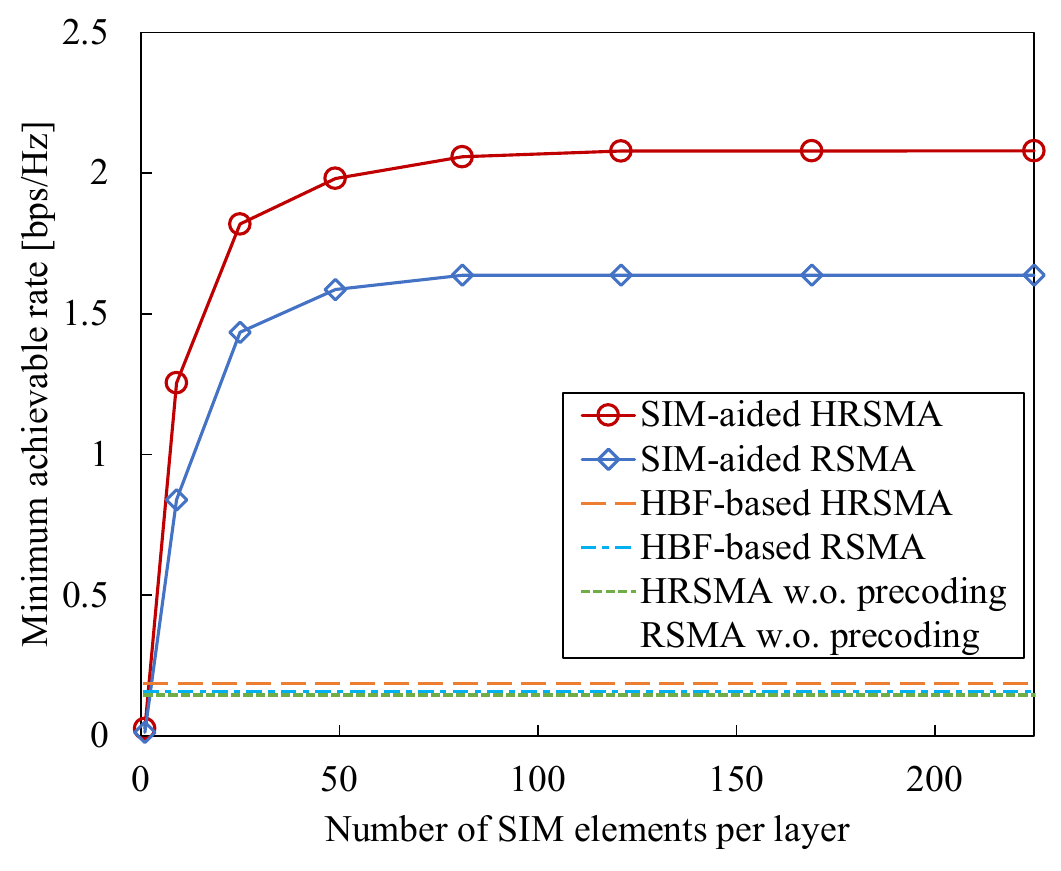}
    \caption{Minimum achievable rate versus the number of SIM elements per layer.}
    \label{fig:num_element}
\end{figure}

Fig.~\ref{fig:num_element} investigates the impact of the number of SIM elements per layer on the minimum achievable rate. 
As the number of elements increases, the minimum rate rapidly improves initially, then gradually saturates, indicating that additional meta-atoms enhance wavefront shaping and interference suppression up to a certain limit. 
This trend demonstrates that the SIM architecture achieves scalable analog beamforming capability even under massive-array configurations.
When the number of SIM elements is small, the SIM aperture facilitates only coarse spatial sampling of the incident field, resulting in limited controllability of the reflected and transmitted phases. 
As the number of SIM elements increases, the spatial sampling density improves, allowing the SIM to synthesize more accurate and finely shaped wavefronts that can redirect energy more effectively toward the desired users. 
This enhances both inter-user interference mitigation and signal focusing, leading to a rapid growth in the minimum user rate.
However, the rate improvement eventually saturates beyond a threshold number of elements. 
Since the element spacing is fixed, increasing the number of SIM elements enlarges the total aperture area rather than the spatial sampling density.
As the aperture becomes sufficiently large, it covers the entire effective far-field beam footprint, and additional elements observe nearly identical electromagnetic fields.
Consequently, adjacent elements become highly spatially correlated, and no new independent radiation modes can be generated.
In this regime, further enlargement of the SIM aperture provides negligible additional wavefront control capability, leading to saturation of the achievable minimum rate.
Consequently, additional meta-atoms introduce redundant phase control without expanding the effective channel rank or improving the achievable rate.
Across all configurations, SIM-aided HRSMA achieves the highest minimum rate, confirming that the combination of hierarchical rate-splitting and wave-domain channel reconfiguration maximizes both fairness and throughput. 
In contrast, conventional HBF and non-precoding schemes exhibit nearly flat and significantly lower performance, since they lack the large-aperture wave-domain gain provided by the SIM.

\begin{figure}[t]
    \centering
    \includegraphics[width=0.95\hsize]{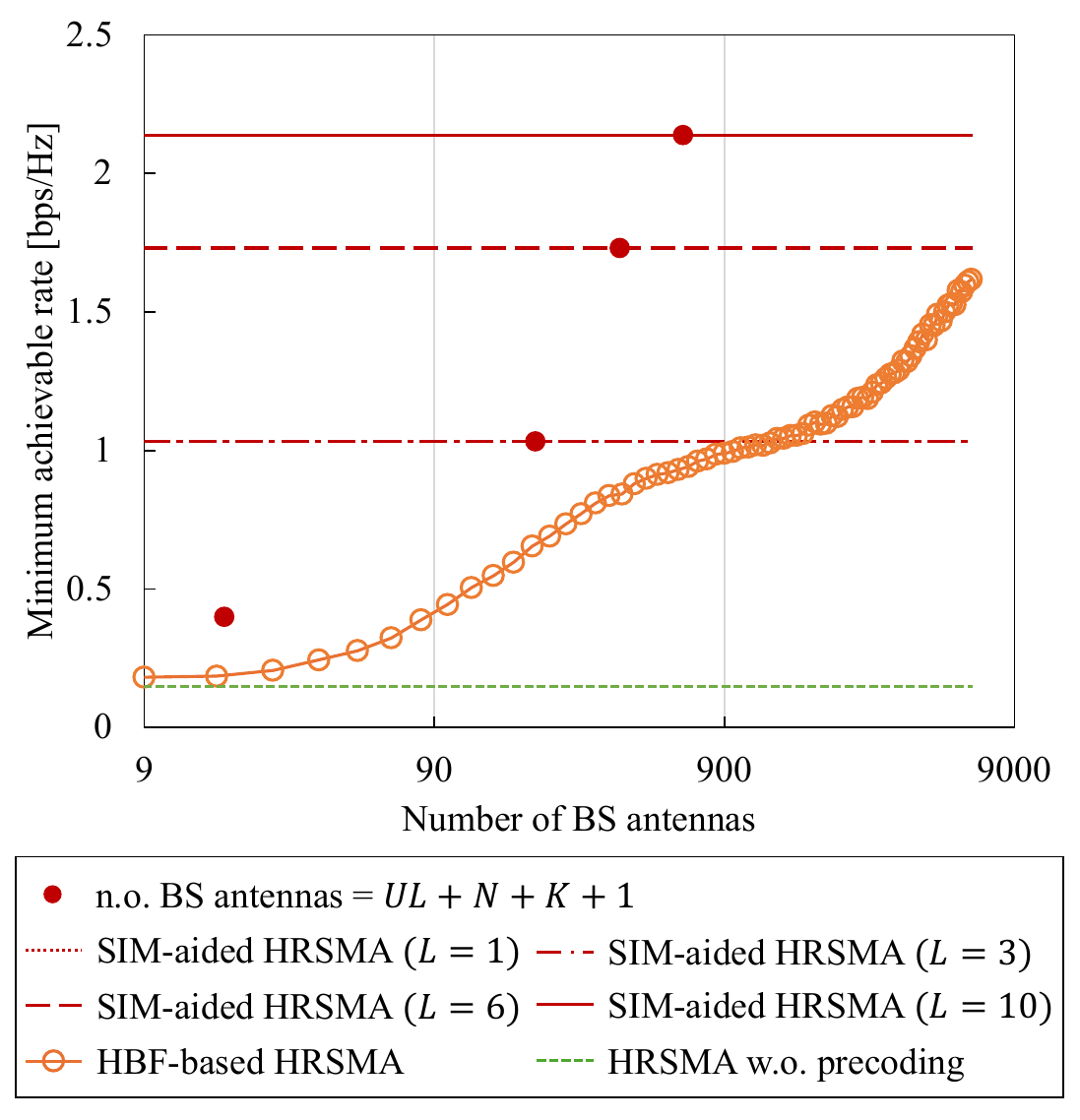}
    \caption{Minimum achievable rate versus the number of BS antennas.}
    \label{fig:num_BS_antenna}
\end{figure}

Fig.~\ref{fig:num_BS_antenna} compares the minimum achievable rate of SIM-aided HRSMA with that of HBF-based HRSMA as a function of the number of BS antennas.
For the HBF-based scheme, the performance monotonically improves as the number of active antennas increases, reflecting the enhancement of the spatial degrees of freedom available for analog beamforming.
In contrast, SIM-aided HRSMA attains comparable or even higher performance with significantly fewer active BS antennas, as the SIM layers effectively contribute additional wave-domain degree of freedom via passive reconfiguration.
The red markers indicate the points where the total number of SIM elements and BS antennas in the SIM-aided HRSMA equals the total number of antennas in the HBF-based scheme, i.e., $N_{\text{t}} = U L + N + K + 1$.
These points clearly demonstrate that even when the overall number of radiating and reflecting elements is identical, SIM-aided HRSMA achieves substantially higher minimum rates compared to HBF-based HRSMA.
This indicates that the spatial degrees of freedom introduced by SIM are more efficiently utilized than those generated by simply increasing the number of transmit antennas.
It should be noted that the SIM elements and BS antennas are not hardware-equivalent entities.
Each BS antenna actively generates an independent RF signal, whereas each SIM element passively
modulates the incident wavefront. Nevertheless, comparing the total number of antennas and SIM elements
provides a reasonable measure of the system's spatial degrees of freedom, as both contribute to the 
effective aperture and wave manipulation capability. 
HBF-based systems rely on digital precoding to weight existing channel directions, whereas SIM-aided HRSMA physically reconfigures the propagation operator through multi-layer wavefront manipulation, providing richer and more orthogonal effective channels.
Consequently, the SIM architecture achieves high fairness and robustness with fewer active RF chains, highlighting its potential as a low-cost and energy-efficient alternative to massive antenna arrays.

\begin{figure}[t]
    \centering
    \includegraphics[width=0.95\hsize]{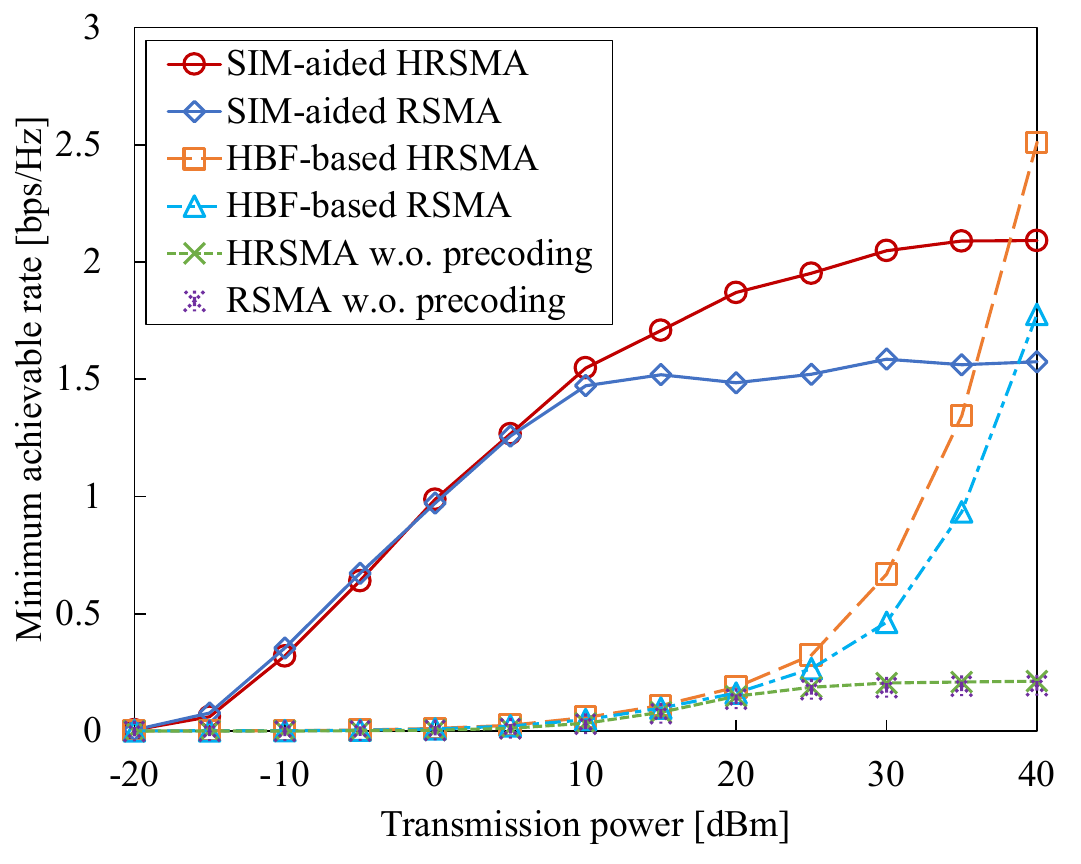}
    \caption{Minimum achievable rate versus transmission power.}
    \label{fig:transmission_power}
\end{figure}

Fig.~\ref{fig:transmission_power} shows the minimum achievable rate as a function of the BS transmit power. 
When the transmission power is low ($P_{\mathrm{t}} < 5$ dBm), such that the SNR is low, SIM-aided HRSMA and SIM-aided RSMA significantly outperform the hybrid beamforming (HBF)-based and non-precoding counterparts. 
However, HBF-based schemes remain nearly flat in this region because the baseband precoders cannot exploit meaningful channel directions when the instantaneous received SNR is low. 
Although the analog beamforming stage provides limited array gain, with a limited number of antenna elements, the hybrid structure cannot reshape the propagation environment.
In contrast, the SIM acts as a large reconfigurable aperture that modifies the effective channel $\mathbf{H}_{\mathrm{eff}}$ by coherently focusing energy toward each user.  
Hence, the SIM-aided systems efficiently convert transmit power into transmission rate, achieving a clear advantage in the low-SNR regime.
In the range of $P_\text{t} \approx 5$-$25$ dBm when the transmission power is in the middle range, SIM-aided HRSMA achieves the highest minimum rate and exhibits a larger slope than the other schemes. 
The improvement arises from two complementary mechanisms: the multi-layer SIM provides additional controllable propagation modes, which increase the smallest singular value of $\mathbf{H}_{\mathrm{eff}}$ and improves fairness among users; and the hierarchical rate-splitting structure redistributes group-common and private streams, alleviating inter-group interference and ensuring that the weakest user benefits from higher energy focusing.  
The SIM-aided RSMA follows a similar trend but lacks the inter-group splitting gain, resulting in a slightly smaller minimum rate.  
HBF-based schemes exhibit improvement with increasing power because the digital precoders can now effectively separate users, although they remain constrained by hybrid hardware and a fixed propagation environment.
When the transmission power is high ($P_{\mathrm{t}} > 25$ dBm), the SIM-aided curves gradually saturate. This saturation occurs because the SIM, while capable of reshaping the channel, cannot increase its effective rank indefinitely.  
Once the main propagation modes are exploited, further power increases only amplify the already dominant links, and the minimum user rate is limited by residual interference and channel correlation.  
In contrast, HBF-based HRSMA and RSMA continue to grow slowly at high power levels, as digital precoding can further suppress inter-user interference in this interference-limited region.  
The non-precoding baselines, however, remain nearly constant across all power levels because they lack any interference mitigation or aperture control.
Overall, the results demonstrate that the proposed SIM-aided HRSMA architecture achieves superior energy efficiency and fairness in the low- to mid-SNR regime, where conventional HBF and non-precoding systems fail to fully exploit the available power.  
By reconfiguring the wavefront in the physical domain and jointly optimizing message hierarchies, the SIM-aided system effectively bridges the gap between analog wave control and logical rate-splitting coordination.

\begin{figure}[t]
    \centering
    \includegraphics[width=0.95\hsize]{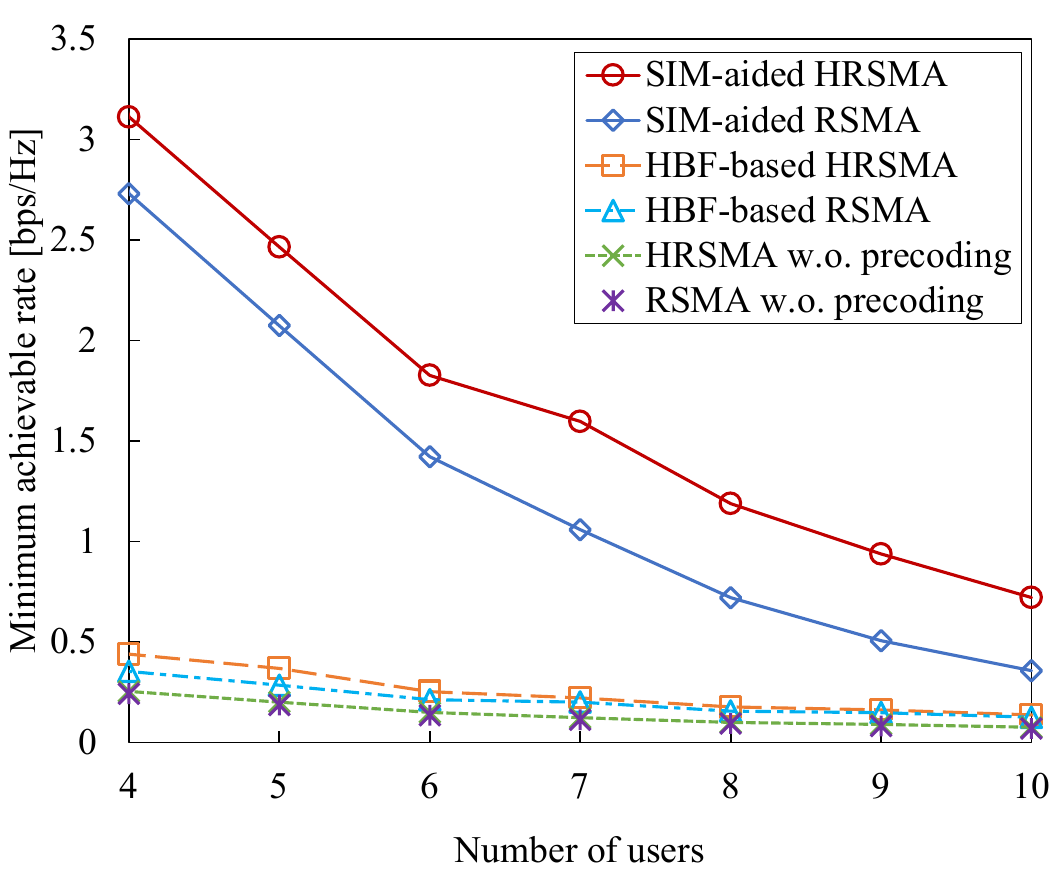}
    \caption{Minimum achievable rate versus the number of users.}
    \label{fig:num_users}
\end{figure}

Fig.~\ref{fig:num_users} illustrates the minimum achievable rate as a function of the number of users $K$. 
The minimum user rate decreases with increasing $K$ for all schemes, because more users introduce additional inter-user interference and reduce the available transmit power and degrees of freedom per user.  
However, the rate degradation differs significantly across architectures.
The proposed SIM-aided HRSMA consistently achieves the highest minimum rate across all user configurations. 
This robustness arises from the combination of wave-domain channel reconfiguration and hierarchical rate-splitting. 
SIM dynamically redistributes wave energy to balance received power among spatially separated users, 
while the HRSMA structure allows fine-grained management of inter-group and intra-group interference.  
As $K$ increases, these mechanisms jointly maintain higher fairness, preventing severe performance degradation of the bottleneck user.
SIM-aided RSMA also outperforms all HBF-based and non-precoding schemes; however, it exhibits slightly lower minimum rates than SIM-HRSMA. 
This difference originates from the absence of group-common messages in RSMA, 
which limits its ability to compensate for users experiencing poor channel conditions.
In contrast, HBF-based HRSMA and RSMA yield relatively low, since their digital precoders can only manage interference in the limited baseband dimension, 
without the benefit of wave-domain control.  
Their fairness capability is fundamentally constrained by the number of RF chains and by the inability to reshape the propagation environment.  
The non-precoding baselines remain nearly flat and lowest among all schemes, 
indicating that without any precoding or SIM-assisted focusing, the weakest user's rate is almost unaffected by increasing $K$, because all users experience similarly from power and interference limitations.
Overall, these results demonstrate that even under dense-user scenarios, SIM provides sufficient spatial adaptability to redistribute energy and preserve fairness, whereas conventional HBF architectures cannot compensate for the increased user coupling. 
The hierarchical structure further enhances resilience against user growth, confirming the scalability of the proposed system design.

\begin{figure}[t]
    \centering
    \includegraphics[width=0.95\hsize]{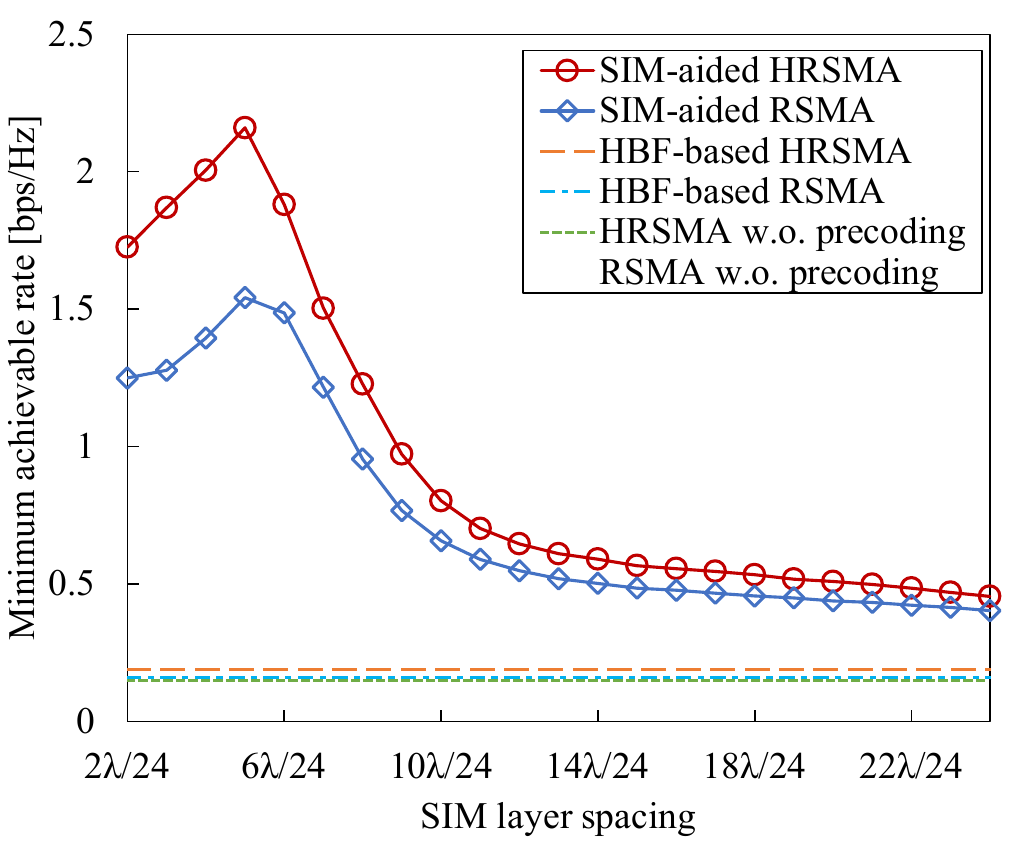}
    \caption{Minimum achievable rate versus SIM layer spacing.}
    \label{fig:layer_spacing}
\end{figure}

Fig.~\ref{fig:layer_spacing} depicts the minimum achievable rate versus the SIM inter-layer spacing $d_\ell$.
Both SIM-aided HRSMA and SIM-aided RSMA exhibit a clear peaking behavior: the rate initially increases as $d_\ell$ grows from very small values, reaches a maximum at approximately $d_\ell \!\approx\! 5\lambda/24$, and then gradually decreases as $d_\ell$ becomes larger. 
The non-monotonic trend can be explained by Rayleigh-Sommerfeld diffraction theory between adjacent layers, whose complex gain contains an amplitude factor proportional to $\cos\eta/t$ and a phase factor $e^{j2\pi t/\lambda}$, where $t$ is the inter-element distance and $\eta$ is the incidence angle.
When $d_\ell$ is too small, rays connecting an element to the next layer tend to impinge at large obliquity angles, so that the $\cos\eta$ factor decreases. 
Therefore, the effective inter-layer propagation weakens, and the end-to-end path loss is large, which results in a lower capacity.
As $d_\ell$ increases to moderate values, the product of distance loss $1/t$ and obliquity $\cos\eta$ improves.
Moreover, the accumulated phase rotation across layers is sufficiently diverse to realize fine wavefront shaping; this yields the observed performance peak.
When $d_\ell$ becomes too large, phase rotation between an element and its counterparts in the next layer varies slowly across neighboring elements.
Consequently, the inter-element phase gradient available for wave manipulation is reduced, limiting the SIM's ability to finely steer or focus the electromagnetic field.
Consequently, the controllable spatial degrees of freedom do not increase further, and the minimum rate degrades.

\section{Discussion}
\label{section:discussion}

The simulation results presented in the previous section comprehensively verify that the proposed SIM-aided HRSMA architecture improves both spectral efficiency and user fairness while drastically reducing reliance on digital beamforming resources.
This section discusses the physical insights, implementation trade-offs, and potential limitations of the proposed design.

\subsection{Effectiveness of SIM-Aided HRSMA}

According to the aforementioned performance evaluation, the proposed SIM-aided HRSMA outperforms both conventional HBF-based and non-precoding baselines.
This superiority stems from the integration of logical-layer interference management via hierarchical rate-splitting and wave-domain cascaded control via SIM.
Whereas traditional HBF architectures perform beamforming by linearly combining antenna outputs based on channel coefficients,
SIM reconstructs the effective channel operator through multilayer wavefront shaping.
Each meta-atom performs a progressive phase transformation on the incident wave,
achieving simultaneous energy focusing toward weak users and interference suppression.
By integrating physical-layer wave reconfiguration with the logical-layer hierarchical decoding structure,
the proposed system effectively manages both inter-group and intra-group interference without digital precoding.

\subsection{Design Trade-off: Increasing HBF Antennas vs. Introducing SIM Layers}

From a practical design perspective, it is essential to examine how the system should be enhanced by increasing the number of antennas in HBF or by integrating SIM layers.
In HBF, increasing the number of antennas enlarges the active aperture and improves array gain,
but it requires additional phase shifters and active antennas, leading to higher hardware cost, power consumption, and calibration complexity.
Moreover, the rank of the channel matrix is physically constrained by the scattering environment,
so simply adding more antennas does not generate new independent propagation paths.

In contrast, SIM offers a more energy-efficient and scalable alternative.
Each SIM layer functions as a passive or semi-passive aperture that excites and redistributes reflection and diffraction paths,
thereby expanding the set of available spatial modes.
Thus, SIM can increase the effective channel rank without additional RF hardware.
However, this increase in degrees of freedom is bounded by physical constraints such as aperture size, inter-layer distance, and operating frequency band.
At the system level, these characteristics make SIM particularly advantageous in low-to-medium SNR or weakly scattering environments,
where conventional antenna densification is less effective.

\subsection{Cost and Complexity Implications of Replacing Digital Precoding with SIM}

Replacing digital precoding with physical-layer wavefront control yields specific benefits in terms of hardware, computational, and operational efficiency.

\subsubsection{Hardware Cost and Energy Efficiency}
Digital or hybrid beamforming architectures require high-power baseband processors.
In contrast, the SIM front-end performs spatial signal processing in the wave domain using passive or semi-passive meta-atom arrays.
Each element is controlled by only a low-power bias circuit (e.g., varactor diodes or PIN switches),
so even with hundreds of elements, the total control power remains on the order of milliwatts.
Only a few active antennas are needed at the base station, dramatically reducing hardware cost and energy consumption.

\subsubsection{Computational and Optimization Complexity}
Digital precoding requires high-dimensional linear algebra operations to derive the precoding matrix.
Matrix inversion and singular value decomposition operations have cubic complexity with respect to the number of data streams and must be updated for each channel realization within the channel coherence time.
In frequency-selective channels, this process must be repeated for every subcarrier,
imposing a heavy computational burden and latency constraints on the baseband unit.
Hybrid beamforming further increases complexity by requiring joint analog-digital optimization, often involving iterative convex solvers.

In contrast, the SIM-aided architecture eliminates digital matrix-based precoding.
Spatial control is achieved through nonlinear wavefront manipulation,
and optimization can be performed via low-dimensional stochastic methods such as SPSA.
This algorithm estimates the gradient direction using only two function evaluations per iteration,
regardless of the number of variables.
Consequently, the computational complexity scales linearly with the number of meta-atoms rather than cubically with the number of data streams,
making the SIM approach far more suitable for large-scale deployments.

\subsubsection{Operational and Maintenance Cost}
HBF requires frequent calibration to compensate for hardware imperfections such as phase noise and mutual coupling.
Since SIM primarily consists of passive components, it does not require calibration at the same frequency.
However, periodic calibration is still necessary to account for element variation and temperature drift.
Furthermore, multilayer and large-scale element control introduces constraints on feedback bandwidth and update delay,
which may cause performance degradation in highly dynamic environments.

Overall, replacing digital precoding with SIM represents a paradigm shift
from computationally-intensive linear signal manipulation to energy-efficient spatial reconfiguration in the wave domain.
In practice, a hybrid design combining SIM-based wavefront control with lightweight digital assistance using a small number of RF chains
is expected to achieve an optimal balance between performance, cost, and flexibility.
Such an approach positions the SIM-aided architecture as a promising physical-layer foundation
for future 6G networks that demand high capacity, strong fairness, and sustainability.

\section{Conclusion}
\label{section:conclusion}
In this study, a novel transmission paradigm that integrates HRSMA with SIMs was investigated, enabling nonlinear wave-domain signal processing without relying on digital precoding. 
Using the proposed alternating optimization framework, which jointly refines SIM phase configurations, transmit power, and user grouping, the system achieves a balanced tradeoff between spectral efficiency and user fairness.
Simulation results demonstrate that the proposed SIM-aided HRSMA system consistently outperforms HBF-based and non-precoding systems in terms of minimum rate, spectral efficiency, and fairness.  
Beyond the numerical gains, the results reveal fundamental insights into the physical and architectural behavior of SIM-assisted transmission.  
The nonlinear multi-layer interactions within the SIM form controllable propagation operators that extend the spatial degrees of freedom beyond linear precoding, enabling effective interference redistribution and fairness improvement even under dense-user and power-limited conditions.  
Furthermore, analysis of the design parameters indicates that performance saturates with excessive layering or the number of elements, suggesting the existence of an optimal physical configuration that maximizes field diversity while minimizing spatial correlation.  
Finally, the comparison with HBF architectures demonstrates that SIM achieves superior efficiency per controllable unit, delivering higher minimum rates even with the same number of electromagnetic elements.  
From a practical standpoint, replacing digital precoding with SIM-based wavefront control simplifies the transceiver architecture, reduces the number of active RF chains, and mitigates baseband computation, pointing toward an energy-efficient and scalable design paradigm for beyond-6G systems.

Future research will focus on extending SIM-aided HRSMA to wideband and multi-cell environments, investigating real-time adaptation of metasurface phases under imperfect channel knowledge, and developing hybrid SIM and digital architectures that dynamically balance flexibility and complexity.

\section*{Acknowledgment}
A portion of the results presented in this paper was obtained through research commissioned by JSPS KAKENHI Grant Number 24K23850.

\bibliographystyle{IEEEtran}
\bibliography{reference}

\begin{thebibliography}{10}
\providecommand{\url}[1]{#1}
\csname url@samestyle\endcsname
\providecommand{\newblock}{\relax}
\providecommand{\bibinfo}[2]{#2}
\providecommand{\BIBentrySTDinterwordspacing}{\spaceskip=0pt\relax}
\providecommand{\BIBentryALTinterwordstretchfactor}{4}
\providecommand{\BIBentryALTinterwordspacing}{\spaceskip=\fontdimen2\font plus
\BIBentryALTinterwordstretchfactor\fontdimen3\font minus \fontdimen4\font\relax}
\providecommand{\BIBforeignlanguage}[2]{{%
\expandafter\ifx\csname l@#1\endcsname\relax
\typeout{** WARNING: IEEEtran.bst: No hyphenation pattern has been}%
\typeout{** loaded for the language `#1'. Using the pattern for}%
\typeout{** the default language instead.}%
\else
\language=\csname l@#1\endcsname
\fi
#2}}
\providecommand{\BIBdecl}{\relax}
\BIBdecl

\bibitem{IoT}
A.~Al-Fuqaha, M.~Guizani, M.~Mohammadi, M.~Aledhari, and M.~Ayyash, ``Internet of things: A survey on enabling technologies, protocols, and applications,'' \emph{IEEE Communications Surveys \& Tutorials}, vol.~17, no.~4, pp. 2347--2376, 2015.

\bibitem{IoTJ_1}
L.~D. Xu, W.~He, and S.~Li, ``Internet of things in industries: A survey,'' \emph{IEEE Transactions on Industrial Informatics}, vol.~10, no.~4, pp. 2233--2243, 2014.

\bibitem{mina}
M.~Kato, T.~Koketsu~Rodrigues, T.~Abe, and T.~Suganuma, ``Exploiting radio frequency characteristics with a support unmanned aerial vehicle to improve wireless sensor location estimation accuracy,'' \emph{IEEE Internet of Things Journal}, vol.~11, no.~24, pp. 39\,570--39\,578, 2024.

\bibitem{6G_resource_management}
Z.~Zhang, Y.~Xiao, Z.~Ma, M.~Xiao, Z.~Ding, X.~Lei, G.~K. Karagiannidis, and P.~Fan, ``{6G} wireless networks: Vision, requirements, architecture, and key technologies,'' \emph{IEEE Vehicular Technology Magazine}, vol.~14, no.~3, pp. 28--41, 2019.

\bibitem{Boya_HDMA_JSAC}
R.~Deng, B.~Di, H.~Zhang, and L.~Song, ``{HDMA}: Holographic-pattern division multiple access,'' \emph{IEEE Journal on Selected Areas in Communications}, vol.~40, no.~4, pp. 1317--1332, 2022.

\bibitem{OMA_NOMA}
M.~Zeng, A.~Yadav, O.~A. Dobre, G.~I. Tsiropoulos, and H.~V. Poor, ``Capacity comparison between {MIMO-NOMA} and {MIMO-OMA} with multiple users in a cluster,'' \emph{IEEE Journal on Selected Areas in Communications}, vol.~35, no.~10, pp. 2413--2424, 2017.

\bibitem{NOMA_1}
Z.~Ding, X.~Lei, G.~K. Karagiannidis, R.~Schober, J.~Yuan, and V.~K. Bhargava, ``A survey on non-orthogonal multiple access for {5G} networks: Research challenges and future trends,'' \emph{IEEE Journal on Selected Areas in Communications}, vol.~35, no.~10, pp. 2181--2195, 2017.

\bibitem{NOMA_2}
Z.~Ding, Z.~Yang, P.~Fan, and H.~V. Poor, ``On the performance of non-orthogonal multiple access in {5G} systems with randomly deployed users,'' \emph{IEEE Signal Processing Letters}, vol.~21, no.~12, pp. 1501--1505, 2014.

\bibitem{RSMA_2}
Y.~Mao, O.~Dizdar, B.~Clerckx, R.~Schober, P.~Popovski, and H.~V. Poor, ``Rate-splitting multiple access: Fundamentals, survey, and future research trends,'' \emph{IEEE Communications Surveys \& Tutorials}, vol.~24, no.~4, pp. 2073--2126, 2022.

\bibitem{RSMA_4}
H.~Joudeh and B.~Clerckx, ``Rate-splitting for max-min fair multigroup multicast beamforming in overloaded systems,'' \emph{IEEE Transactions on Wireless Communications}, vol.~16, no.~11, pp. 7276--7289, 2017.

\bibitem{HRSMA_1}
J.~Zhou, Y.~Sun, Q.~Cao, and C.~Tellambura, ``Total delay optimization in cache-enabled {C-RANs} with hierarchical rate splitting,'' \emph{IEEE Transactions on Vehicular Technology}, vol.~71, no.~11, pp. 11\,832--11\,846, 2022.

\bibitem{HRSMA_2}
Z.~Li, C.~Ye, Y.~Cui, S.~Yang, and S.~Shamai, ``Rate splitting for multi-antenna downlink: Precoder design and practical implementation,'' \emph{IEEE Journal on Selected Areas in Communications}, vol.~38, no.~8, pp. 1910--1924, 2020.

\bibitem{HRSMA_3}
J.~Yang, S.~Gao, Y.~Lu, Z.~Yang, and G.~Tu, ``Max-min fairness of cooperative rsma for aggregated {VLC/RF} systems,'' \emph{IEEE Communications Letters}, vol.~29, no.~3, pp. 552--556, 2025.

\bibitem{RSMA_1}
B.~Clerckx, Y.~Mao, E.~A. Jorswieck, J.~Yuan, D.~J. Love, E.~Erkip, and D.~Niyato, ``A primer on rate-splitting multiple access: Tutorial, myths, and frequently asked questions,'' \emph{IEEE Journal on Selected Areas in Communications}, vol.~41, no.~5, pp. 1265--1308, 2023.

\bibitem{RSMA_beamforming}
O.~Dizdar, Y.~Mao, and B.~Clerckx, ``Rate-splitting multiple access to mitigate the curse of mobility in (massive) {MIMO} networks,'' \emph{IEEE Transactions on Communications}, vol.~69, no.~10, pp. 6765--6780, 2021.

\bibitem{Boya_IRS_TVT}
H.~Zhang, B.~Di, L.~Song, and Z.~Han, ``Reconfigurable intelligent surfaces assisted communications with limited phase shifts: How many phase shifts are enough?'' \emph{IEEE Transactions on Vehicular Technology}, vol.~69, no.~4, pp. 4498--4502, 2020.

\bibitem{NZhang_1}
S.~Mao, C.~Yuen, L.~Liu, M.~Xiao, S.~Yu, and N.~Zhang, ``{RIS}-enhanced semantic-aware sensing, communication, computation and control for internet of things,'' \emph{IEEE Transactions on Wireless Communications}, pp. 1--1, 2025.

\bibitem{Hashida_PartialBlockage_TCCN}
H.~Hashida, Y.~Kawamoto, and N.~Kato, ``Mathematical modeling and deployment optimization: Intelligent reflecting surface-aided communications under partial blockages,'' \emph{IEEE Transactions on Cognitive Communications and Networking}, vol.~11, no.~5, pp. 3306--3316, 2025.

\bibitem{SIM_1}
J.~An, C.~Yuen, C.~Xu, H.~Li, D.~W.~K. Ng, M.~Di~Renzo, M.~Debbah, and L.~Hanzo, ``Stacked intelligent metasurface-aided {MIMO} transceiver design,'' \emph{IEEE Wireless Communications}, vol.~31, no.~4, pp. 123--131, 2024.

\bibitem{RSMA_3}
B.~Clerckx, Y.~Mao, R.~Schober, and H.~V. Poor, ``Rate-splitting unifying {SDMA, OMA, NOMA}, and multicasting in {MISO} broadcast channel: A simple two-user rate analysis,'' \emph{IEEE Wireless Communications Letters}, vol.~9, no.~3, pp. 349--353, 2020.

\bibitem{Hashida_FluidAntenna_JSAC}
S.~Zhang, Y.~Zhang, H.~Hashida, Y.~C. Eldar, M.~D. Renzo, and B.~Di, ``Fluid antenna systems enabled by reconfigurable holographic surfaces: Beamforming design and experimental validation,'' \emph{IEEE Journal on Selected Areas in Communications}, pp. 1--1, 2025.

\bibitem{IRS_1}
C.~Huang, A.~Zappone, G.~C. Alexandropoulos, M.~Debbah, and C.~Yuen, ``Reconfigurable intelligent surfaces for energy efficiency in wireless communication,'' \emph{IEEE Transactions on Wireless Communications}, vol.~18, no.~8, pp. 4157--4170, 2019.

\bibitem{NZhang_2}
S.~Mao, N.~Zhang, J.~Hu, K.~Yang, Y.~Xiong, and X.~Chen, ``Computation bits maximization for {IRS}-aided mobile-edge computing networks with phase errors and transceiver hardware impairments,'' \emph{IEEE Transactions on Vehicular Technology}, vol.~73, no.~4, pp. 5587--5601, 2024.

\bibitem{Hibi_near_TCCN}
R.~Hibi, H.~Hashida, Y.~Kawamoto, and N.~Kato, ``Angle–distance separated hierarchical beam training with multi-focal and defocal beams for large-scale {IRS}-aided near-field communication,'' \emph{IEEE Transactions on Cognitive Communications and Networking}, pp. 1--1, 2025.

\bibitem{Boya_IRSNOMA_NetworkM}
S.~Fu, Y.~Wang, X.~Feng, B.~Di, and C.~Li, ``Reconfigurable intelligent surface assisted non-orthogonal multiple access network based on machine learning approaches,'' \emph{IEEE Network}, vol.~38, no.~2, pp. 272--279, 2024.

\bibitem{IRS_2}
E.~Bj\"ornson, {\"O}.~{\"O}zdogan, and E.~G. Larsson, ``Intelligent reflecting surface versus decode-and-forward: How large surfaces are needed to beat relaying?'' \emph{IEEE Wireless Communications Letters}, vol.~9, no.~2, pp. 244--248, 2020.

\bibitem{Hashida_sharing_WCM}
H.~Hashida, Y.~Kawamoto, and N.~Kato, ``Sharing intelligent reflecting surfaces in multi-operator communication systems for sustainable {6G} networks,'' \emph{IEEE Wireless Communications}, pp. 1--8, 2025.

\bibitem{IRS_3}
Z.~Wang, L.~Liu, and S.~Cui, ``Channel estimation for intelligent reflecting surface assisted multiuser communications: Framework, algorithms, and analysis,'' \emph{IEEE Transactions on Wireless Communications}, vol.~19, no.~10, pp. 6607--6620, 2020.

\bibitem{Hashida_Sharing_TCCN}
H.~Hashida, Y.~Kawamoto, and N.~Kato, ``Machine learning-based infrastructure sharing and shared operations for intelligent reflecting surface-aided communications,'' \emph{IEEE Transactions on Cognitive Communications and Networking}, vol.~10, no.~1, pp. 198--208, 2024.

\bibitem{IRS_RSMA_1}
A.~A. Salem, S.~Abdallah, M.~Saad, K.~Alnajjar, and M.~A. Albreem, ``Robust secure {ISAC}: How {RSMA} and active {RIS} manage eavesdropper's spatial uncertainty,'' \emph{IEEE Transactions on Vehicular Technology}, pp. 1--16, 2025.

\bibitem{IRS_RSMA_2}
I.~Aboumahmoud, E.~Hossain, and A.~Mezghani, ``Resource management in {RIS}-assisted rate splitting multiple access for next generation {(xG)} wireless communications: Models, state-of-the-art, and future directions,'' \emph{IEEE Communications Surveys \& Tutorials}, vol.~27, no.~3, pp. 1618--1655, 2025.

\bibitem{IRS_RSMA_3}
S.~Singh, K.~Singh, S.~K. Singh, H.~Shin, and T.~Q. Duong, ``Multiple access for holographic reconfigurable intelligent surface {(HRIS)}-aided near-field communications,'' \emph{IEEE Internet of Things Journal}, pp. 1--1, 2025.

\bibitem{IRS_HRSMA_1}
T.~Qi, B.~Lyu, and C.~You, ``Transmit power minimization for {IRS}-assisted hierarchical rate-splitting multiple access systems,'' \emph{IEEE Wireless Communications Letters}, vol.~14, no.~5, pp. 1386--1390, 2025.

\bibitem{IRS_HRSMA_2}
Z.~Yang, L.~Feng, F.~Zhou, K.~Xie, X.~Qiu, Z.~Xiong, C.~Yuen, and Z.~Han, ``User grouping-based hierarchical rate-splitting multiple access for reconfigurable intelligent surface-assisted downlink communication,'' \emph{IEEE Transactions on Cognitive Communications and Networking}, pp. 1--1, 2025.

\bibitem{IRS_HRSMA_3}
A.~Bansal, K.~Singh, and C.-P. Li, ``Analysis of hierarchical rate splitting for intelligent reflecting surfaces-aided downlink multiuser {MISO} communications,'' \emph{IEEE Open Journal of the Communications Society}, vol.~2, pp. 785--798, 2021.

\bibitem{Hibi_Near_WCM}
R.~Hibi, H.~Hashida, Y.~Kawamoto, and N.~Kato, ``{IRS}-aided near-field communication: Prospects and challenges with codebook approach,'' \emph{IEEE Wireless Communications}, vol.~32, no.~4, pp. 156--161, 2025.

\bibitem{SIM_2}
A.~Papazafeiropoulos, J.~An, P.~Kourtessis, T.~Ratnarajah, and S.~Chatzinotas, ``Achievable rate optimization for stacked intelligent metasurface-assisted holographic {MIMO} communications,'' \emph{IEEE Transactions on Wireless Communications}, vol.~23, no.~10, pp. 13\,173--13\,186, 2024.

\bibitem{SIM_machine_learning}
G.~Huang, J.~An, Z.~Yang, L.~Gan, M.~Bennis, and M.~Debbah, ``Stacked intelligent metasurfaces for task-oriented semantic communications,'' \emph{IEEE Wireless Communications Letters}, vol.~14, no.~2, pp. 310--314, 2025.

\bibitem{SIM_RSMA}
Q.~Huai, Y.~Liang, and W.~Yuan, ``Stacked intelligent metasurfaces-aided rate splitting multiple access system,'' \emph{IEEE Wireless Communications Letters}, vol.~14, no.~7, pp. 2224--2228, 2025.

\bibitem{SIM_JSAC}
J.~An, C.~Xu, D.~W.~K. Ng, G.~C. Alexandropoulos, C.~Huang, C.~Yuen, and L.~Hanzo, ``Stacked intelligent metasurfaces for efficient holographic mimo communications in {6G},'' \emph{IEEE Journal on Selected Areas in Communications}, vol.~41, no.~8, pp. 2380--2396, 2023.

\bibitem{SPSA}
J.~Spall, ``Multivariate stochastic approximation using a simultaneous perturbation gradient approximation,'' \emph{IEEE Transactions on Automatic Control}, vol.~37, no.~3, pp. 332--341, 1992.

\end{thebibliography}

\begin{IEEEbiography}[{\includegraphics[width=1in,height=1.25in,clip,keepaspectratio]{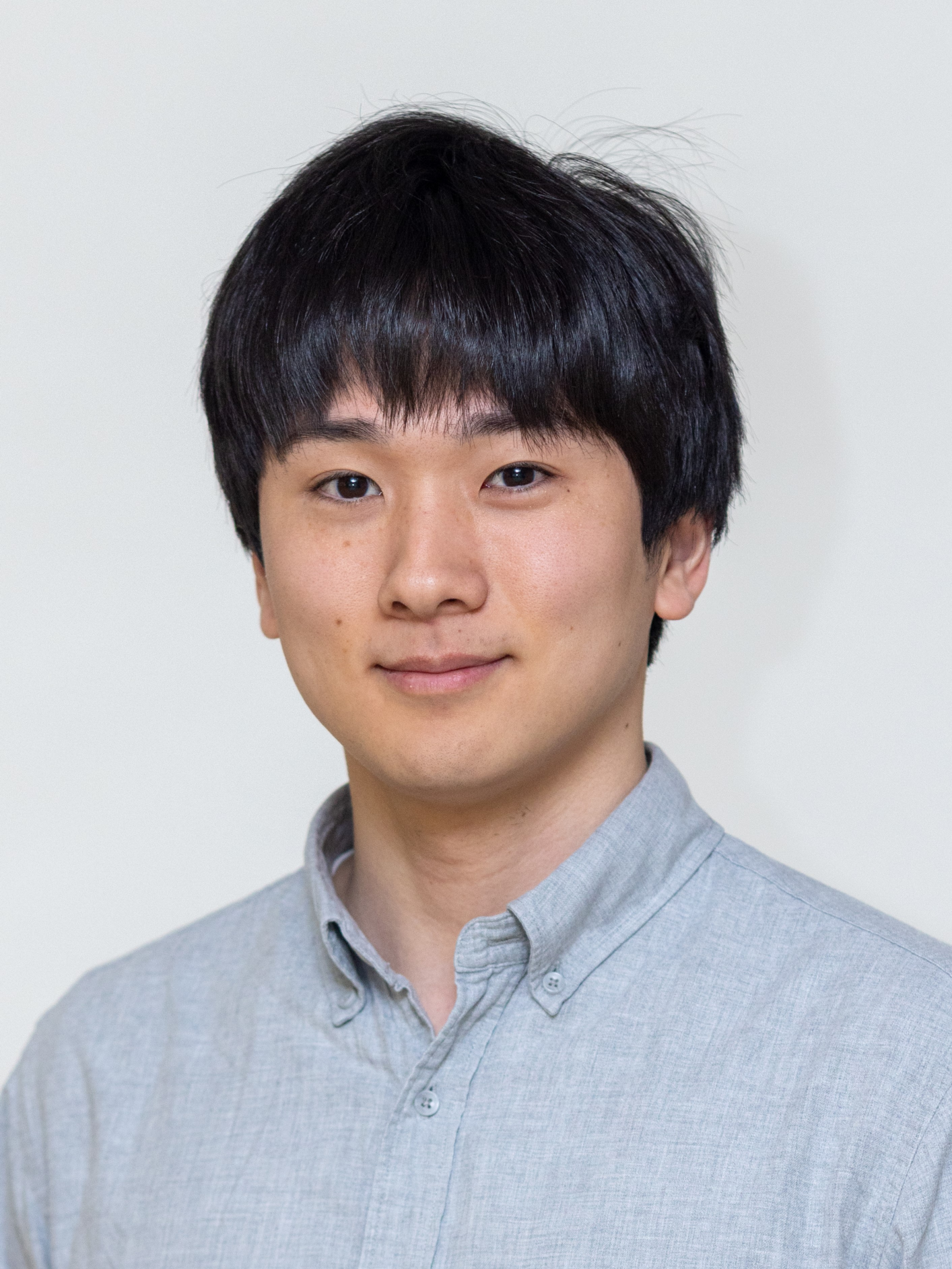}}]
{Hiroaki Hashida}
(M'24) is an Assistant Professor with the Frontier Research Institute for Interdisciplinary Sciences, Tohoku University, Japan. He received his Ph.D. in 2024 from Tohoku University, Sendai, Japan. He was a recipient of the Presidential Award for Outstanding Students from Tohoku University and the Ikushi Prize from the Japan Society for the Promotion of Science in 2024. His research interests include wireless communication networks and intelligent surface-aided wireless communication systems. He is a member of the IEEE and the Institute of Electronics, Information, and Communication Engineers.
\end{IEEEbiography}

\begin{IEEEbiography}[{\includegraphics[width=1in,height=1.25in,clip,keepaspectratio]{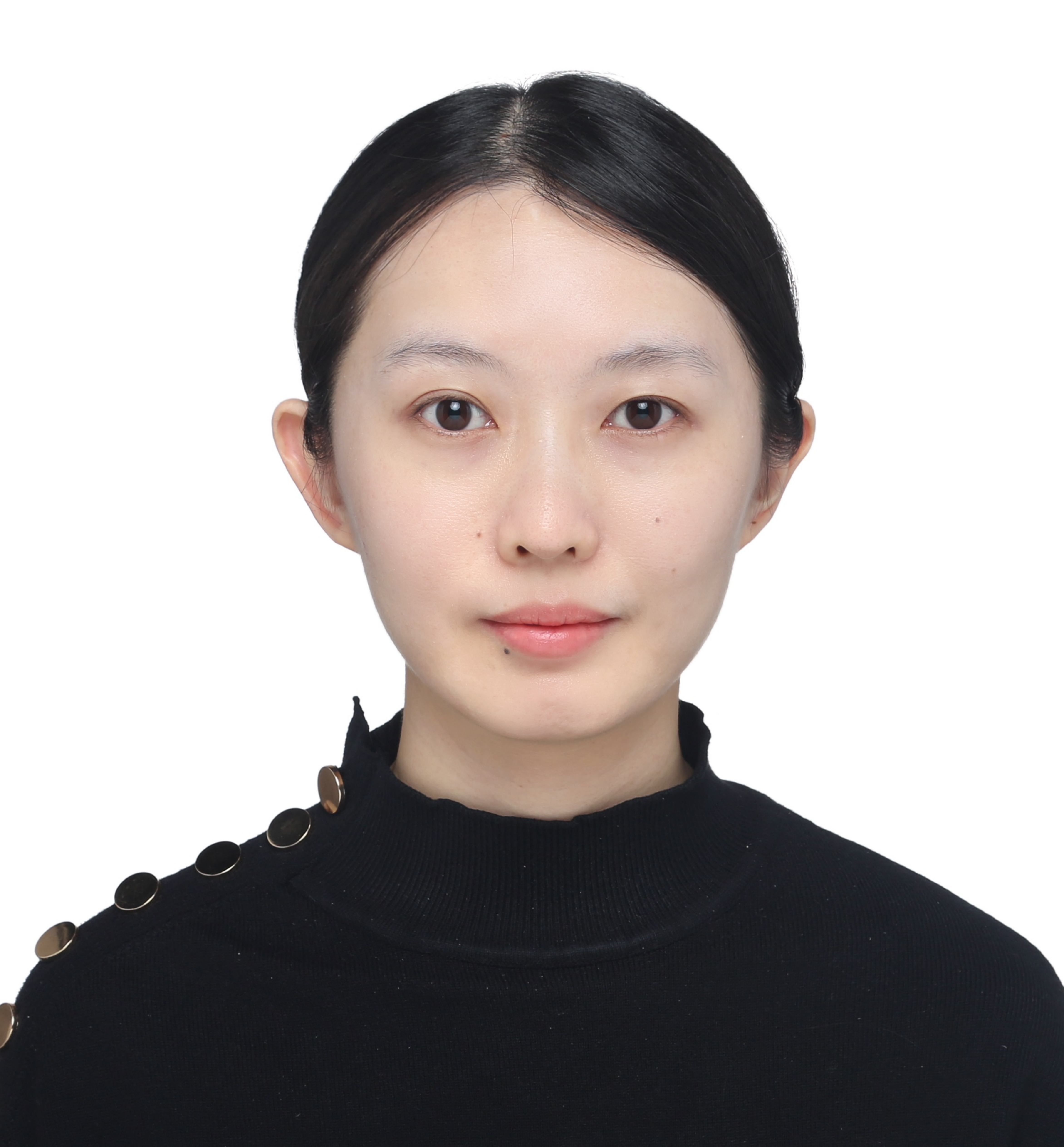}}]
{Boya Di}
(SM'24) received the Ph.D. degree from the Department of Electronics, Peking University, China, in 2019. She was a Post-Doctoral Researcher at Imperial College London. She has been an Assistant Professor with Peking University since 2021. Her current research interests include holographic surfaces, AI-enabled communications, and aerial access networks. She serves as an Associate Editor for IEEE Transactions on Vehicular Technology, IEEE Communications Surveys and Tutorials, and IEEE Internet of Things Journal.
\end{IEEEbiography}

\end{document}